# An empirical formulation of accelerated molecular dynamics for simulating and predicting microstructure evolution in materials

Liang Wan [a,b,*], Qingsong Mei [c], Haowen Liu [d,*], Huafeng Zhang [a,b], Jun-Ping Du [e], Shigenobu Ogata [e], Wen Tong Geng [f,*]

[a] *School of Physics and Optoelectronic Engineering, Yangtze University, Jingzhou 434023, China*

[b] *Key Laboratory of Micro-Nano Photonic Materials and Devices, Yangtze University, Jingzhou 434023, China*

[c] *School of Power and Mechanical Engineering, Wuhan University, Wuhan 430072, China*

[d] *School of Computer Science, Wuhan University, Wuhan 430072, China*

[e] *Department of Mechanical Science and Bioengineering, The University of Osaka, Osaka 560-8531, Japan*

[f] *Department of Physics, Zhejiang Normal University, Jinhua 321004, China*

[*] Corresponding authors.

*E-mail addresses:* lwan5@outlook.com (L. Wan), hwenliu@whu.edu.cn (H. Liu), wtgeng@zjnu.edu.cn (W.T. Geng).

## ABSTRACT

Despite its widespread use in materials science, conventional molecular dynamics (MD) simulations are severely constrained by timescale limitations. To address this shortcoming, we propose an empirical formulation of accelerated MD method, adapted from a collective-variable-based extended system dynamics framework. While this framework is originally developed for efficient free energy sampling and reaction pathway determination of specific rare events in condensed matter, we have modified it to enable accelerated MD simulation and prediction of microstructure evolution of materials across a broad range of scenarios. In essence, the nearest neighbor off-centering absolute displacement (NNOAD), which quantifies the deviation of an atom from the geometric center of its nearest neighbors in materials, is introduced. We propose that the collection of NNOADs of all atoms can serve as a generalized reaction coordinate for various structural transitions in materials. The NNOAD of each atom, represented by its three components, is coupled with three additional dynamic variables assigned to the atom. Time evolution of the additional dynamic variables follows Langevin equation, while Nosé-Hoover dynamics is employed to thermostat the system. Through careful analysis and benchmark simulations, we established appropriate parameter ranges for the equations in our method. Application of this method to several test cases demonstrates its capability to accelerate MD simulations by several orders of magnitude while maintaining kinetic consistency and good accuracy in predicting long timescale microstructure evolutions of materials. We also provide some preliminary thoughts on theoretical justification of the method, offering insights into its underlying principles.

**Keywords:** accelerated molecular dynamics; extended system dynamics; dynamic equations; generalized reaction coordinate; atomic shuffling motion; microstructure evolution

## 1. Introduction

Molecular dynamics (MD) simulation is widely applicated in condensed matter physics and materials science [1,2]. By solving the equations of motion which do not explicitly account for quantum effects (e.g., Newton's equations) for particles (atoms or molecules) under a predefined force field, MD simulations generate detailed phase space trajectories of the system. These trajectories provide profound physical insights into the behavior and properties of the system under study [3,4]. Despite its versatility, the conventional MD method is fundamentally constrained by its limited timescale. This restriction poses significant challenges for investigating many critical microstructure evolution processes in materials, which often occur over timescales far beyond the reach of standard MD simulations [5].

To address this challenge, various accelerated MD methods have been developed [6-16]. Basically, it is the structural transition events rather than the atomic vibrations that is of interest in plenty of studies. By examining the potential energy surface (PES) of the simulated system, one can see that the time spent in atomic vibrations within the basins of PES depends strongly on heights of saddle points (i.e., the activation barriers) between these basins, and it is the crossing of these saddles leads to the occurrence of structural transitions. A bias potential can thus be made to reduce the effective height of these barriers, enabling faster transitions. Methods which implement this bias-potential-based scheme include hyperdynamics [6,10,13,15], metadynamics [9,11,12], and the adaptive boost method [14]. These methods differ primarily in their formulation of the bias potential and treatment of saddle crossings.

The bias potential is often expressed as a function of one or multiple collective variables (CVs), which are typically formulated based on atomic coordinates [17]. A well-chosen set of CVs should accurately describe the reaction paths of the structural transitions of interest [18-20]. However, identifying optimal CVs is challenging due to the diversity of structural transitions in condensed matter systems and the lack of prior knowledge about their reaction paths [17,19,20]. Common practice involves tailoring CVs to the specific problem, such as using bond lengths for atomic diffusion [10] or local atomic strain for dislocation dynamics [13]. The reliability of bias-potential-based methods critically depends on the choice of CVs, as poor selections can lead to unreliable mechanics and kinetics due to issues like 'hidden barriers' [21].

A significant advancement in accelerated MD methods is the temperature accelerated molecular dynamics (TAMD) approach [22], a variant of adiabatic free energy dynamics [23,24]. Unlike bias-potential-based methods, TAMD introduces extra dynamic variables corresponding to CVs, couples them harmonically to the CVs, and accelerates sampling of microstates along the CV-constrained paths by applying an artificially high temperature to these extra dynamic variables [22]. This approach allows for an increased number of CVs without significantly sacrificing efficiency [25,26] and aligns with the extended system dynamics philosophy, a milestone in MD method development [5,27-32].

However, TAMD (as well as some other methods like metadynamics [9,12]) was primarily designed for free energy landscape sampling or determining reaction pathways of specific rare events. For materials scientists, a key objective of MD simulations lies in predicting the state-to-state processes of microstructure evolution, which typically consist of a series of diverse structural transition events. Therefore, there is a need to go beyond the current formulations of TAMD.

In this work, we propose an empirical formulation termed 'shuffling accelerated molecular dynamics (SAMD)'. After detailing the SAMD method, we demonstrate its application to a benchmark problem to determine appropriate parameter values. Preliminary validation through several case studies highlights its effectiveness and consistency in accelerating simulations and predicting microstructure evolution. We conclude with a brief theoretical analysis and qualitative justification of the method.

## 2. Formulation of the SAMD method

### 2.1. Definition of nearest neighbor off-centering absolute displacement (NNOAD) of atom

As mentioned earlier, most accelerated MD methods require a proper formulation of collective variables (CVs) to describe the reaction paths of structural transition events. However, the atomic motions involved in these transitions can vary significantly. For example, vacancy or interstitial migration in a crystal lattice involves directional atomic movement accompanied by minor lattice distortion, while dislocation nucleation or glide typically involves local shear transformations and non-affine atomic motions. Similarly, martensitic transformations often require both coordinated ('military') and irregular ('civilian') atomic motions [33,34].

To describe such irregular or non-affine atomic motions, the concept of 'shuffling motion' of atoms has been widely adopted in materials science [33,35-41]. Analysis of various structural transitions suggests that atomic shuffling motion is a common feature of thermally activated events during microstructure evolution. This generality motivates us to use atomic shuffling motion as a basis for formulating a generalized reaction coordinate applicable to a wide range of structural transitions in materials.

However, there is no consensus on a formal definition of shuffling motion of atoms. A formulation has been proposed based on non-affine transformations within a local atomic neighborhood [42,43]. Albeit physically reasonable, this formulation is computationally cumbersome, requiring frequent updates of a reference configuration. Here we propose an alternative definition using a quantity called the nearest neighbor off-centering absolute displacement (NNOAD). The NNOAD of atom $i$, denoted as $\boldsymbol{d}_i$, is defined as follows and illustrated in Fig. 1(a):

$$\boldsymbol{d}_i \equiv (d_i^{\mathrm{X}},\, d_i^{\mathrm{Y}},\, d_i^{\mathrm{Z}}) = \left(\left|R_i^{\mathrm{X}}(\boldsymbol{X})\right|,\, \left|R_i^{\mathrm{Y}}(\boldsymbol{X})\right|,\, \left|R_i^{\mathrm{Z}}(\boldsymbol{X})\right|\right) \;,$$

$$\boldsymbol{R}_i(\boldsymbol{X}) \equiv \left(R_i^{\mathrm{X}}(\boldsymbol{X}),\, R_i^{\mathrm{Y}}(\boldsymbol{X}),\, R_i^{\mathrm{Z}}(\boldsymbol{X})\right) = \boldsymbol{x}_i - \frac{1}{N_i(r_D)} \sum_{j \in \tilde{N}_i(r_D)} \boldsymbol{x}_j \;, \tag{1}$$

where $d_i^{\mathrm{X}}, d_i^{\mathrm{Y}}, d_i^{\mathrm{Z}}$ are the components of $\boldsymbol{d}_i$ along the X, Y, Z axes respectively, $\boldsymbol{X} = (\boldsymbol{x}_1, \boldsymbol{x}_2, \dots \boldsymbol{x}_N)$ is the collection of Cartesian coordinates of all atoms with $\boldsymbol{x}_i = (x_i^{\mathrm{X}}, x_i^{\mathrm{Y}}, x_i^{\mathrm{Z}})$, $\boldsymbol{R}_i(\boldsymbol{X})$ stands for the displacement of atom $i$ relative to the geometric center of its nearest neighbors as illustrated in Fig. 1(a), $R_i^{\mathrm{X}}(\boldsymbol{X})$, $R_i^{\mathrm{Y}}(\boldsymbol{X})$, $R_i^{\mathrm{Z}}(\boldsymbol{X})$ represent the components of $\boldsymbol{R}_i(\boldsymbol{X})$ along the X, Y, Z axes respectively, $N_i(r_D)$ is the number of neighboring atoms within a spherical cutoff distance $r_D$, and $\widetilde{N}_i(r_D)$ is the set of these neighbors.

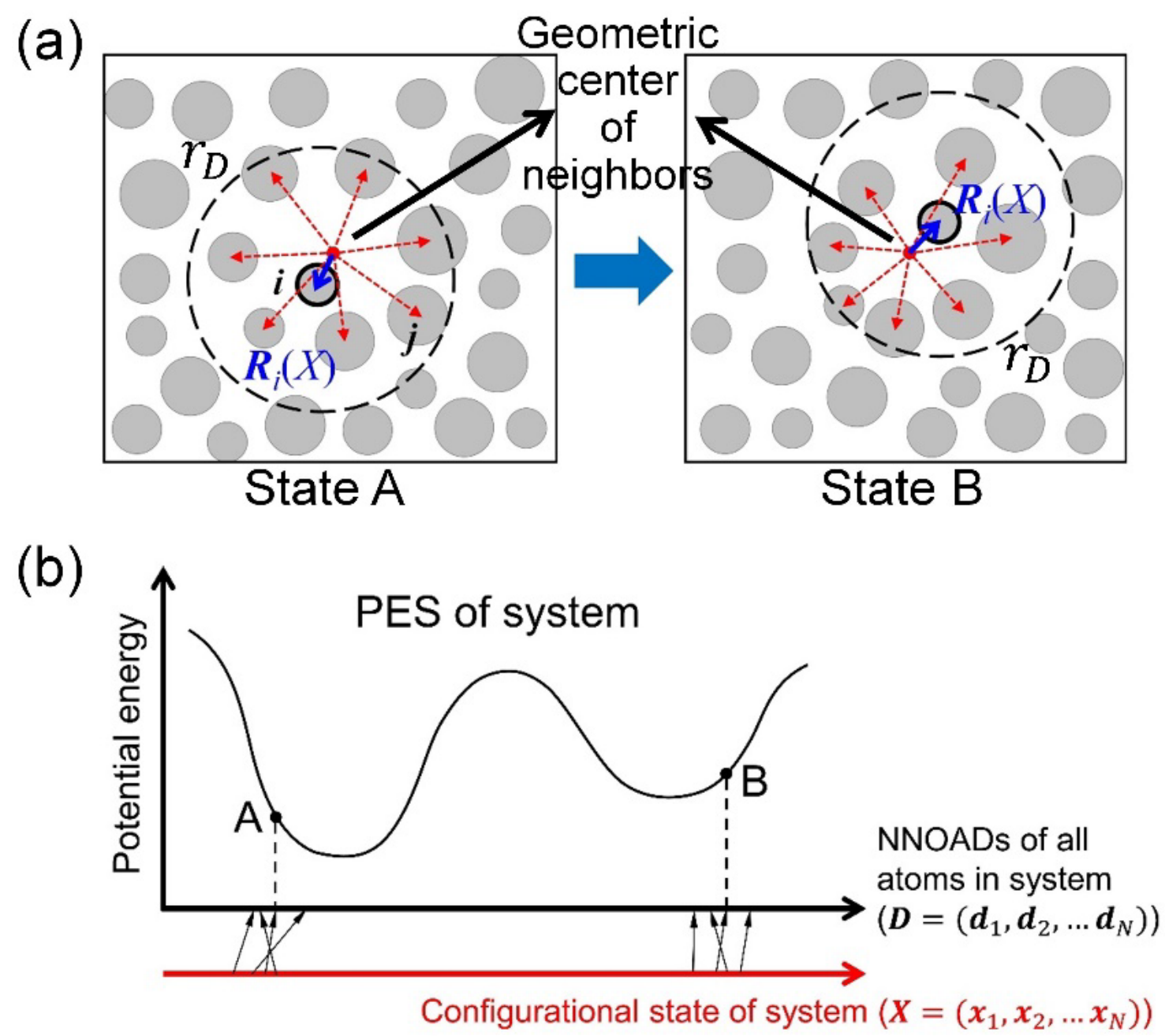

**Fig. 1.** (a) Schematic illustration of definition and calculation of NNOAD for a single atom in a condensed matter system, shown in configurational state A and B. Grey spheres of varying sizes represent atoms of different species. The atom of interest (labelled $i$) is highlighted with a black circle, while its neighbors within the spherical cutoff distance $r_D$ are labelled $j$. The geometric center of these neighboring atoms is marked by red dots. (b) Simplified two-dimensional illustration of PES for the system in (a). The positions of states A and B on the PES are indicated. The abscissa represents the NNOADs of all atoms in the system ($\boldsymbol{D} = (\boldsymbol{d}_1, \boldsymbol{d}_2, \dots \boldsymbol{d}_N)$), with a mapping of individual configurational states in Cartesian coordinates ($\boldsymbol{X} = (\boldsymbol{x}_1, \boldsymbol{x}_2, \dots \boldsymbol{x}_N)$) to the NNOADs shown at the bottom.

The NNOAD is a non-negative 3-dimensional vector that quantifies the deviation of an atom from the geometric center of its nearest neighbors, effectively characterizing its relative motion with respect to the closest surrounding atoms. We propose that, with an appropriate choice of $r_D$ (e.g., the first valley in the material's radial distribution function (RDF)), the NNOAD provides a good measure of atomic shuffling motion in materials.

For any structural transition in materials, the NNOADs of all atoms are expected to exhibit specific variations and one-to-one mapping with configurational changes along the reaction path, as schematically illustrated in Figs. 1(a) and 1(b). Although the precise relationship between NNOAD values and configurational state changes is unknown and can be complex, we propose that the collection of NNOADs of all atoms can serve as a generalized reaction coordinate for describing the reaction paths of diverse structural transitions. Additionally, the computational simplicity of NNOAD, as shown in Eq. (1), offers significant advantages for practical implementation.

### 2.2. Nosé-Hoover thermostat plus Langevin equation for dynamics of an extended system

The TAMD method introduced the framework of an extended system with extra dynamic variables harmonically coupled to CVs [22]. Inspired by this framework, we propose an accelerated MD method for simulating and predicting microstructure evolution of materials under given thermodynamic conditions. Specifically, the use of CVs is eliminated. Instead, we couple the NNOADs of all atoms to the extra dynamic variables in dynamic equations.

In the TAMD method, different kinds of dynamic equations — such as Langevin dynamics [44,45] and Nosé-Hoover dynamics [29-31] — can be used to describe atomic motions and the time evolution of extra dynamic variables, provided they sample the system's microstates in a canonical distribution [22]. After extensive trial simulations, we found that combining Nosé-Hoover dynamics for atomic motions with the Langevin equation for the extra dynamic variables yields optimal results. The equations of motion for the whole extended system are as follows:

$$-\frac{\partial U_{\kappa}(\boldsymbol{X},\boldsymbol{S})}{\partial x_i^{\alpha}} - \zeta m_i \dot{x}_i^{\alpha} = m_i \ddot{x}_i^{\alpha} , \quad \dot{\zeta} = \frac{\gamma_x^2}{3Nk_BT}\left\{\sum_{i=1}^{N} m_i \dot{\boldsymbol{x}}_i^2 - 3Nk_BT\right\} ,$$

$$-\frac{\partial U_{\kappa}(\boldsymbol{X},\boldsymbol{S})}{\partial s_i^{\alpha}} - \gamma_s m_i \dot{s}_i^{\alpha} + \sqrt{2m_i\gamma_s k_B T_s}\, R^s(t) = m_i \ddot{s}_i^{\alpha} ,$$

$$U_{\kappa}(\boldsymbol{X},\boldsymbol{S}) = U(\boldsymbol{X}) + \frac{1}{2}\kappa \sum_{j=1}^{N} \sum_{\alpha\in\{\mathrm{X,Y,Z}\}} \left(s_j^{\alpha} - d_j^{\alpha}(\boldsymbol{X})\right)^2 . \qquad (2)$$

Here, $i \in \{1,2,3,\dots,N\}$ labels each atom in the system, $\alpha \in \{\mathrm{X,Y,Z}\}$ represents the axis of Cartesian coordinates, $m_i$ is the mass of atom $i$, $\boldsymbol{X} = (\boldsymbol{x}_1, \boldsymbol{x}_2, \dots \boldsymbol{x}_N)$ is the collection of atomic coordinates of system with $\boldsymbol{x}_i = (x_i^{\mathrm{X}}, x_i^{\mathrm{Y}}, x_i^{\mathrm{Z}})$, $\boldsymbol{S} = (\boldsymbol{s}_1, \boldsymbol{s}_2, \dots \boldsymbol{s}_N)$ represents the collection of extra dynamic variables of all atoms with $\boldsymbol{s}_i = (s_i^{\mathrm{X}}, s_i^{\mathrm{Y}}, s_i^{\mathrm{Z}})$. The potential $U_\kappa(\boldsymbol{X}, \boldsymbol{S})$ includes the original interatomic potential $U(\boldsymbol{X})$ and a harmonic coupling term between the extra dynamic variables ($s_j^\alpha$) and the NNOADs ($d_j^\alpha(\boldsymbol{X})$), with coupling coefficient κ. $\zeta$ stands for the extra dynamic variable of the thermostat in Nosé-Hoover dynamics. The parameters $\gamma_x$ and $\gamma_s$ are damping coefficients. $T$ and $T_s$ are thermostat temperatures. $R^s(t)$ is the white noise function for Langevin dynamics, and $k_B$ is the Boltzmann constant.

Equations (1) and (2) provide an empirical formulation for accelerated MD simulations of microstructure evolution. This formulation combines Nosé-Hoover dynamics for atomic motions with Langevin dynamics for the extra dynamic variables, harmonically coupling the latter to the NNOADs of the atoms. Notably, the mass of extra dynamic variables of an atom is set equal to the mass of that atom.

In practice, the local arrangement of atoms can change to certain extent so that the nearest neighbors of a specific atom can change on some time step, particularly near saddle points during structural transitions, causing abrupt changes in NNOAD values. These changes can introduce energy pulses and force discontinuities to the physical system due to the harmonic coupling between NNOADs and extra dynamic variables in $U_\kappa(\boldsymbol{X}, \boldsymbol{S})$. To avoid this, we adjust the extra dynamic variables $\boldsymbol{s}_i$ by the same amount as the change in NNOAD for any atom (labeled $i$) whenever a change of its nearest neighbors occurs.

### 2.3. Remarks on our empirical formulation

Our empirical formulation (Eqs. (1) and (2)) for accelerated MD simulation is inspired by and adapted from the TAMD method. However, it differs from TAMD in several key aspects, reflecting its distinct objectives and underlying principles.

**Differences from TAMD**

The TAMD method was primarily developed for efficient calculation of free energy profile or determination of reaction path for a specific rare event (e.g., structural transitions) in chemical physics systems [22,46]. It relies on three conditions: (i) $\gamma_s \gg \gamma_x$, (ii) $\kappa \gg k_B T$, and (iii) $T_s$ being sufficiently large. These conditions ensure adiabatic separation between the dynamics of atomic motions and the extra dynamic variables, enabling accurate free energy calculation and significant acceleration of the rare event in simulation. The escape time $t^{(2)}$ for a structural transition is given by $t^{(2)} = O(\gamma_s \exp(\Delta F/k_B T_s)$, while its original escape time is $t^{(1)} = O(\gamma_x \exp(\Delta F/k_B T))$, where $\Delta F$ is the free energy barrier.

In contrast, our method aims to provide an accelerated MD framework for simulating and predicting microstructure evolutions in various kinds of materials under given thermodynamic conditions. This requires two key features:

(i') **Kinetic Consistency**: The time order and relative frequencies of all the structural transitions predicted in a simulation should align with those from conventional MD simulation under the same thermodynamic conditions (e.g., temperature, stress, strain rate).

(ii') **Extended Time Span**: The physical time span of microstructure evolution simulated (i.e., the sequence of microstates of the system generated) should substantially exceed that achievable by conventional MD at comparable computational cost.

To meet these requirements, our method diverges from TAMD in three critical ways:

**(1) Lifting Adiabaticity**

In TAMD, the adiabatic conditions ($\gamma_s \gg \gamma_x$ and $\kappa \gg k_B T$) result in a speedup ratio $t^{(1)}/t^{(2)} = A(e^{\Delta F})^B$, where $A = O(\gamma_x/\gamma_s)$ and $B = 1/k_B\,(1/T - 1/T_s)$. This nonlinear dependence on $\Delta F$ means transitions with higher barriers are accelerated more than those with lower barriers, leading to inconsistent acceleration. To avoid this, we lift the adiabaticity requirement in our method, enabling a more uniform acceleration effect which we will show later. And one should note that, our method is not designed for free energy calculations.

**(2) Choice of Dynamics**

While TAMD allows the use of any dynamics that samples the canonical distribution [22], our trial simulations (with adiabaticity requirement lifted) reveal significant differences in kinetic behavior depending on the choice of dynamics for $\boldsymbol{X}$ (atomic coordinates) and $\boldsymbol{S}$ (extra dynamic variables). These preliminary simulations indicate that using Nosé-Hoover dynamics for $\boldsymbol{X}$ and Langevin dynamics for $\boldsymbol{S}$ (as in Eq. (2)) yields better consistency of kinetics than other combinations (e.g., Nosé-Hoover for both, Langevin for both, or Langevin for $\boldsymbol{X}$ and Nosé-Hoover for $\boldsymbol{S}$). This suggests that the dynamic characteristics of the equations play a critical role in fulfilling the requirements for an accelerated MD ((i') and (ii') mentioned above). A qualitative analysis of this choice of dynamics will be provided later.

**(3) Use of NNOADs instead of CVs**

Unlike TAMD, which employs a few CVs to coarse-grain the system and calculate free energy profiles with respect to the CVs, our method uses the NNOADs of all atoms (Eq. (1)) as a generalized reaction coordinate. While the NNOADs of all atoms are 3$N$-dimensional (for system of $N$ atoms), they are not CVs in the traditional sense. Instead, they characterize the shuffling motion of atoms and are employed to enhance the apparent frequency of structural transitions through the extended system dynamics, without focusing on free energy calculations. We will later demonstrate that the dynamics of NNOADs are statistically orthogonal to Cartesian vibrational motions of atoms to a significant extent, ensuring robust temperature control of the system via thermostat when enhancing the shuffling motion of atoms by using our method.

**Terminology and implementation**

Given these distinctions — particularly the unique extended system dynamics (Eq. (2)) and the use of NNOADs (Eq. (1)) to characterize the shuffling motion of atoms and serve as a generalized reaction coordinate — we propose naming our method **shuffling accelerated molecular dynamics (SAMD)**. The SAMD method has been implemented within the open-source LAMMPS package [47,48]. Users can readily activate the SAMD functionality by adding just two new commands to their LAMMPS input script. The complete implementation files with example scripts and a documentation are publicly available through the repository link provided in Appendix A.

## 3. Determination of the proper values of parameters in the SAMD method

### 3.1. Proper values of $\delta t$, $T$, $r_D$, and $\gamma_x$

In the SAMD method, the time step $\delta t$ for numerically solving the differential equations of motion (Eq. (2)) can be comparable to or slightly smaller (e.g., half) than that used in conventional MD simulations. The temperature $T$ in Eq. (2), which specifies the target temperature for the Nosé-Hoover thermostat, is typically set to the desired simulation temperature.

Beyond $\delta t$ and $T$, five additional parameters in Eq. (1) and Eq. (2) require tuning: $r_D$, κ, $\gamma_x$, $\gamma_s$ and $T_s$. For $r_D$, the cutoff distance for nearest neighbors, a suitable choice is the location of the first valley in the RDF of the system. If alternative values are used, care must be taken to ensure that the number of nearest neighbors for any atom does not exceed a reasonable upper limit (e.g., 25).

The damping coefficient $\gamma_x$, which governs the relaxation of instantaneous kinetic temperature $T^*$ to the target temperature $T$ in Nosé-Hoover dynamics, is critical for temperature control. The instantaneous kinetic temperature $T^*$ is defined by the average kinetic energy of all atoms in the system [4]

$$T^* = \frac{1}{3Nk_B}\sum_{i=1}^{N} m_i \boldsymbol{v}_i^2 \quad , \tag{3}$$

where $N$ is the total number of atoms, $m_i$ is the mass of atom $i$, $\boldsymbol{v}_i$ is its velocity, and $k_B$ is the Boltzmann constant. In conventional MD simulation, $\gamma_x$ is typically chosen based on the heat conduction or dissipation rate of the system. However, in SAMD simulations, the harmonic coupling between NNOADs and the extra dynamics variables $\boldsymbol{S}$ (Eq. (2)) continuously introduces additional energy into the physical system. To keep $T^*$ close to $T$, a higher value of $\gamma_x$ is required to dissipate this excess energy efficiently. Based on extensive trial simulations, we recommend $\gamma_x$ values in the range of 10.0 ~ 200.0 ps$^{-1}$, with 100.0 ps$^{-1}$ being a suitable choice for most cases.

### 3.2. A benchmark problem for ad hoc determination of proper values of $\kappa$, $\gamma_s$ and $T_s$

In the absence of a theoretical method for determining the values of κ, $\gamma_s$ and $T_s$, a well-designed benchmark problem provides a practical solution. Our benchmark problem consists of two atomistic models of α-Fe crystals: one with 28 carbon interstitials and the other with a single monovacancy, as illustrated in Figs. 2(a) and 2(b), respectively. Both models measure 3.4 nm × 5.2 nm × 5.8 nm, containing 8640 and 8639 Fe atoms, respectively. Common neighbor analysis (CNA) [49] was used to distinguish atoms with body center cubic (BCC) local structure from those without. Interactions between atoms were described using an embedded atom method (EAM) potential, which accurately models the Fe-C binary system [50,51]. The minimum potential energy profiles for carbon interstitial jumps between neighboring octahedral sites and monovacancy migration between BCC Fe lattice sites, calculated using the nudged elastic band (NEB) method [52,53], are shown in Fig. 2(c). The activation barriers are 0.81 eV for carbon interstitial jumps and 0.64 eV for monovacancy migration, consistent with experimental results [50,54,55].

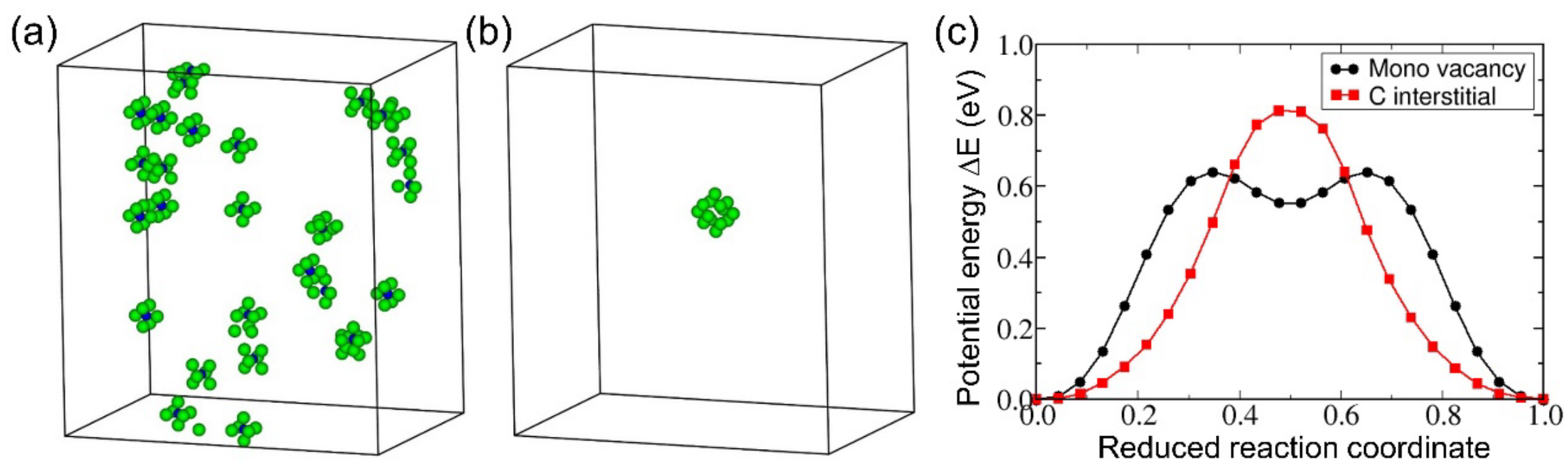

**Fig. 2.** Atomistic models of the benchmark problem. (a) An α-Fe crystal model with 28 carbon interstitials located at the octahedral sites in BCC Fe lattice. (b) An α-Fe crystal model with a single mono vacancy. In both (a) and (b), green spheres represent Fe atoms with non-BCC local structure, while dark blue spheres represent carbon atoms. Fe atoms with BCC local structure are not shown. (c) Minimum potential energy profiles for carbon interstitial jumps between neighboring octahedral sites and monovacancy migration between BCC Fe lattice sites, calculated using the NEB method.

To determine the proper values of κ, $\gamma_s$ and $T_s$, we simulated heat treatment of the α-Fe crystals at $T$ = 750 K using both the SAMD method (with various parameter value sets {κ, $\gamma_s$ , $T_s$ }) and conventional MD. At this temperature, both carbon interstitials and the monovacancy migrate sufficiently to reach a steady state diffusion within computationally accessible timeframes. The apparent migration frequency (Γ) of each defect type was measured in both SAMD and conventional MD simulations. The speedup ratio $\Gamma^{SAMD}/\Gamma^{MD}$, defined as the ratio of migration frequencies in SAMD and conventional MD simulations, quantifies the acceleration effect of the SAMD method. Consistency in acceleration is deemed to be achieved if the speedup ratios for carbon interstitials and monovacancies are nearly identical, provided that there is no kinetic interference from other structural transitions in the simulations. This consistency provides a criterion for evaluating the suitability of the chosen values for the parameters κ, $\gamma_s$, and $T_s$ in SAMD simulations.

The migration distance $R$ of each defect was monitored during simulations. Using random walking theory, the ensemble-averaged square migration distance $\langle R^2 \rangle$ is related to the migration frequency Γ by [56]:

$$\left\langle R^2 \right\rangle = \Gamma \, t \left\langle r^2 \right\rangle \ , \tag{4}$$

where $t$ is the simulation time and $r$ is the jump distance between two successive sites on migration. The speedup ratio is then calculated as:

$$\frac{\Gamma^{SAMD}}{\Gamma^{MD}} = \frac{\left\langle R^2 \right\rangle^{SAMD} / t^{SAMD}}{\left\langle R^2 \right\rangle^{MD} / t^{MD}} \ . \tag{5}$$

Here, $t$ is the product of the time step $\delta t$ and the number of steps $N_{steps}$. Computational overhead from SAMD as compared to conventional MD was negligible and thus ignored.

For carbon interstitials, $\langle R^2 \rangle$ was calculated from the mean square displacement (MSD) of all 28 interstitials. For the mono vacancy, the relationship $\Gamma_{vac} = N_{atoms} \times \Gamma_{Fe}$ was used, where $\Gamma_{vac}$ and $\Gamma_{Fe}$ are the migration frequencies of the mono vacancy and a single Fe atom, respectively, and $N_{atoms}$ is the total number of Fe atoms. Thus, the MSD of the Fe atoms in the monovacancy model was used to calculate $\Gamma_{vac}$.

Fig. 3 shows the root mean square displacement (RMSD) curves for carbon interstitials and equivalent RMSD curves for the monovacancy from conventional MD and SAMD simulations. The equivalent RMSD for the monovacancy is defined as the square root of the sum of squared displacements of all Fe atoms in the model. The values assigned to the simulation parameters, including $\gamma_x$, $T$, κ, $\gamma_S$, and $T_s$, are marked in Fig. 3. No barostat was used, and the models' dimensions were fixed. For SAMD simulations, $r_D$ = 3.46 Å (the first valley in the RDF of ideal BCC Fe) was used, yielding 14 nearest neighbors per Fe atom normally. Time steps were $\delta t$ = 1.0 fs for conventional MD and $\delta t$ = 0.5 fs for SAMD. Three conventional MD simulations (with different initial random atomic velocities) and three SAMD simulations (with different random seeds for Langevin dynamics) for each $T_s$ value (10000 K, 20000 K, 30000 K, 40000 K) were performed. Each conventional MD simulation ran for $2 \times 10^7$ steps, while SAMD simulations ran for $4 \times 10^6$ steps.

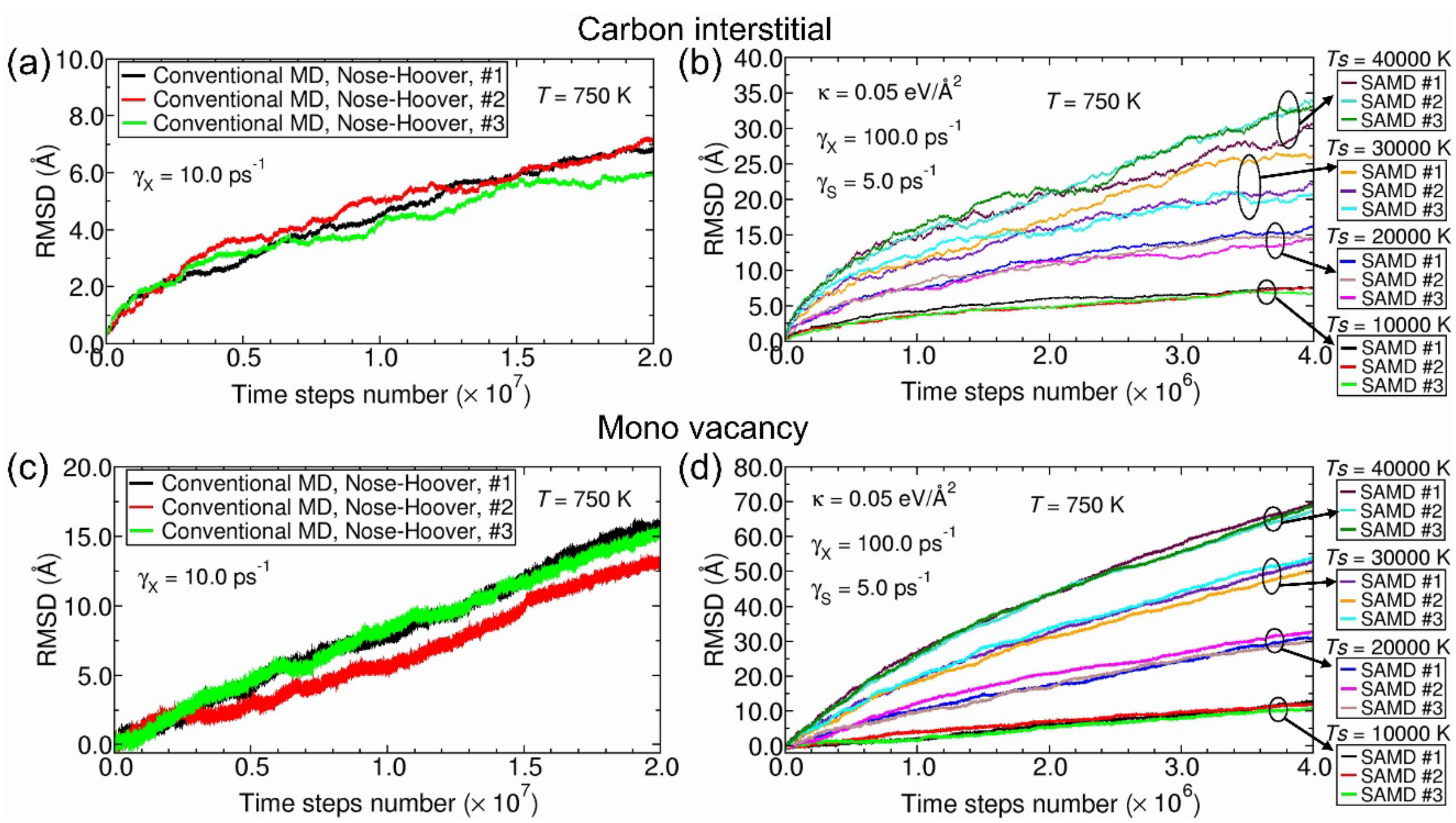

**Fig. 3.** (a) and (b) show the RMSD curves for carbon interstitial migration in α-Fe (model in Fig. 2(a)) from conventional MD and SAMD simulations, respectively. (c) and (d) show the equivalent RMSD curves for monovacancy migration in α-Fe (model in Fig. 2(b)) from conventional MD and SAMD simulations, respectively. The values of simulation parameters are indicated in the graphs.

The RMSD curves in Fig. 3 indicate steady-state diffusion of both defects. Minor deviations among the three curves (#1, #2, #3) for each parameter value set arise from different random values used in simulations. No other structural transitions except the migration of the carbon interstitials or monovacancy were observed, as confirmed by Supplementary Movies S1a and S1b.

Using the same protocol, SAMD simulations with 60 parameter value sets {κ, $\gamma_s$, $T_s$} were performed (see Table 1 for values). For each parameter value set, three SAMD simulations with different random seeds used in Langevin dynamics were conducted, and the MSD of carbon interstitials or Fe atoms was averaged over the three simulations to obtain robust $\langle R^2 \rangle$ values. No other structural transitions were detected. The speedup ratios $\Gamma^{SAMD}/\Gamma^{MD}$ for carbon interstitials ('C') and the monovacancy ('Vac') were thus obtained for all these parameter value sets. In Fig. 4, by plotting $(\Gamma^{SAMD}/\Gamma^{MD})[C]$ on the horizontal axis against $(\Gamma^{SAMD}/\Gamma^{MD})[Vac]$ on the vertical axis, the speedup ratios $\Gamma^{SAMD}/\Gamma^{MD}$ of carbon interstitials and the monovacancy for any given parameter value set of {κ, $\gamma_s$, $T_s$} can be compared. Acceleration is consistent when a data point lies close to the diagonal (i.e., $(\Gamma^{SAMD}/\Gamma^{MD})[C] \approx (\Gamma^{SAMD}/\Gamma^{MD})[Vac]$ ), indicating an appropriate choice of parameter values. Deviations from the diagonal indicate growing inconsistency between the acceleration of the two types of defects for that parameter value set. Based on extensive simulations, suitable ranges for the parameters are:

κ: 0.01 ~ 0.5 eV/Å$^2$

$\gamma_s$: 1.0 ~ 100.0 ps$^{-1}$

The value of $T_s$ should not be excessively large, as it may violate thermal activation condition for structural transitions, leading to ballistic behavior of structural transitions in simulations. Furthermore, we assume the empirically determined ranges for κ and $\gamma_s$ are transferable to SAMD simulations of other materials.

**Table 1**

Values of κ, $\gamma_s$ and $T_s$ used in SAMD simulations with the atomistic models of Figs. 2(a) and 2(b). The total number of combinations of κ, $\gamma_s$ and $T_s$ is 5 × 4 + 6 × 4 + 4 × 4 = 60. Other parameters were fixed: $r_D$ = 3.46 Å, $\delta t$ = 0.5 fs, $T$ = 750 K, $\gamma_x$ = 100 ps$^{-1}$, and $N_{steps}$ = 4 × 10$^6$, consistent with the simulations in Figs. 3(b) and 3(d).

| κ (eV / Å$^2$) | $\gamma_s$ (ps$^{-1}$) | $T_s$ (K) |
|---|---|---|
| 0.05 | 0.1<br>1.0<br>5.0<br>10.0<br>50.0 | 10000<br>20000<br>30000<br>40000 |
| 0.10 | 0.1<br>1.0<br>5.0<br>10.0<br>50.0<br>100.0 | 5000<br>10000<br>15000<br>20000 |
| 0.30 | 1.0<br>5.0<br>10.0<br>50.0 | 2000<br>4000<br>7000<br>9000 |

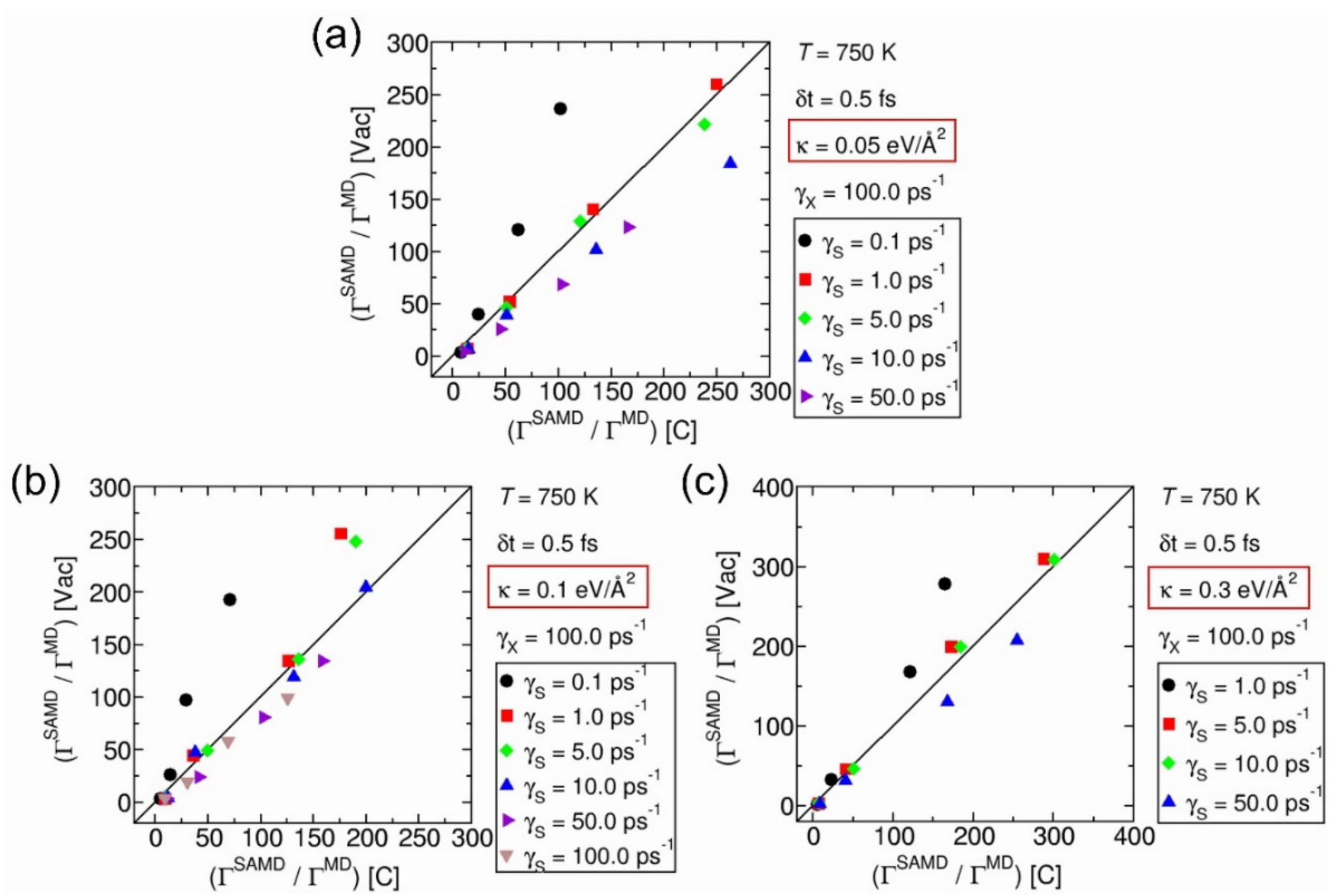

**Fig. 4.** Comparison of the speedup ratio $\Gamma^{SAMD}/\Gamma^{MD}$ for carbon interstitial ('C', horizontal axis) and monovacancy ('Vac', vertical axis) obtained by SAMD simulations performed with different parameter value sets from Table 1, where κ = 0.05 eV / Å$^2$ (a), 0.1 eV / Å$^2$ (b), and 0.3 eV / Å$^2$ (c). The values assigned to simulation parameters of $T$, $\delta t$, κ, $\gamma_x$, and $\gamma_S$ for each data point in the plot are marked aside in each panel.

## 4. Several case studies using the SAMD method

To validate the applicability of the SAMD method for accelerated MD simulations of diverse microstructure evolution behaviors in materials, we selected four representative materials science problems for case studies. In each case, simulations of well-designed atomistic models were performed using both conventional MD and SAMD methods under nearly identical thermodynamic conditions. The SAMD simulation parameters followed the recommendations outlined in the previous section. Results from both methods were carefully analyzed and compared.

### 4.1. Case study (1): Segregation of H atoms on a grain boundary in Al bicrystal

Hydrogen embrittlement poses a significant challenge for metallic engineering materials [57]. The segregation of hydrogen (H) atoms to grain boundaries (GBs) plays a critical role in embrittlement under mechanical loading [58-60]. In this case study, we examine H segregation on Σ5<0 0 1>{3 1 0} symmetric tilt GB in an Al bicrystal. As shown in Fig. 5, the atomistic model contains two identical GBs due to periodic boundary conditions. The model dimensions are 3.2 nm × 3.9 nm × 7.7 nm, comprising 5,664 Al atoms and 195 H atoms initially distributed rather uniformly. An angular-dependent interatomic potential (ADP) [61] was used to describe the Al-H system. This potential accurately predicts properties such as the preference of H atoms for tetrahedral interstitial sites and an energy barrier of 0.189 eV for H migration between tetrahedral and octahedral sites, consistent with first-principles calculations and experiments [61].

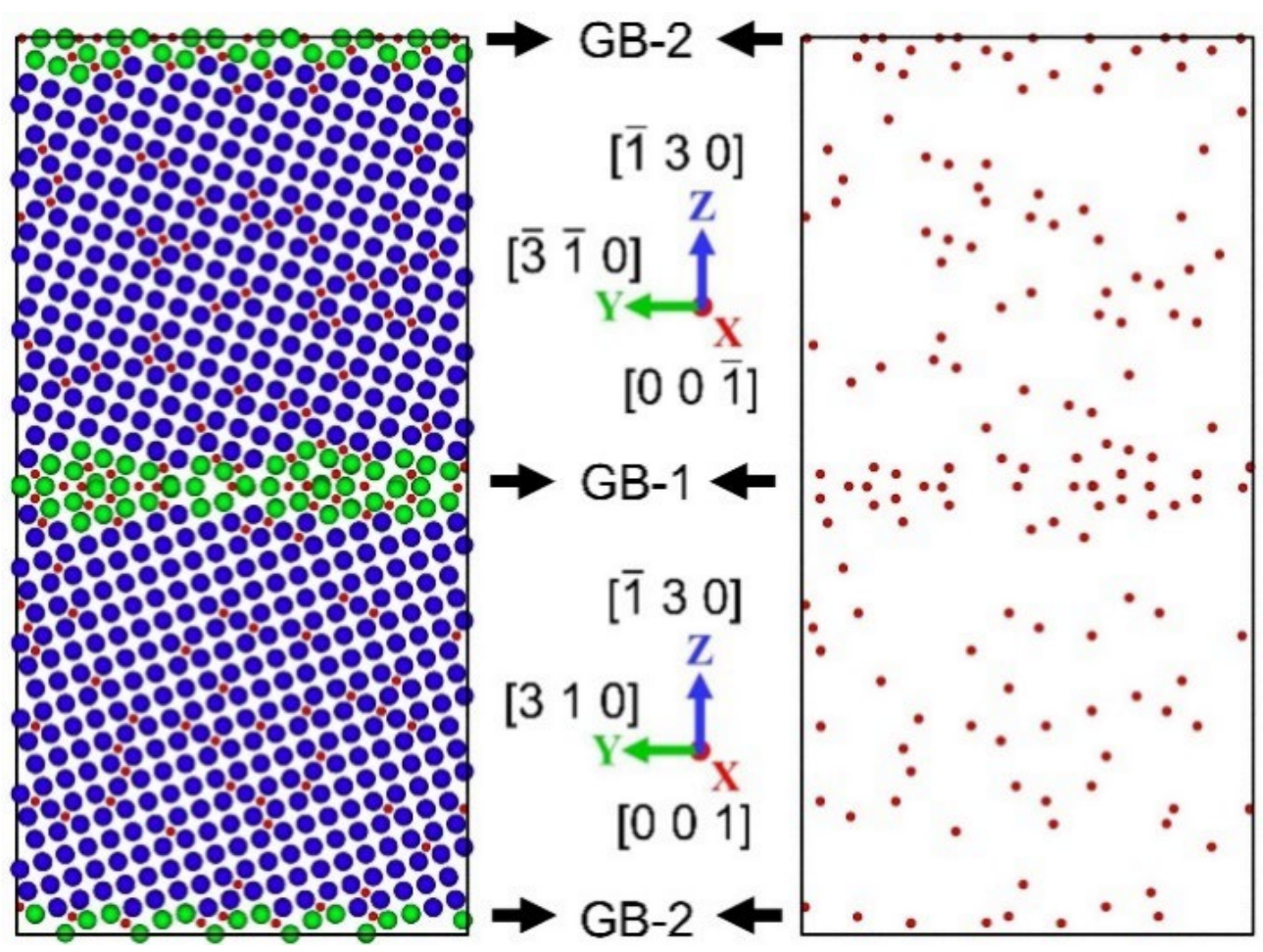

**Fig. 5.** Atomistic model and initial state of an Al bicrystal with the Σ5<001>{310} symmetric tilt GB and dissolved H atoms. Large blue spheres represent Al atoms with FCC local structure, while green spheres represent Al atoms with non-FCC local structure, identified using the CNA method. Small red spheres represent H atoms. Periodic boundary conditions in all three dimensions result in two identical GBs (GB-1 & GB-2). The right panel shows the initial distribution of H atoms.

Conventional MD simulations were performed at $T$ = 100 K and $T$ = 300 K with a time step $\delta t$ = 0.5 fs and $N_{steps}$ = 2 × $10^7$. Temperature control was achieved using the Nosé-Hoover thermostat with $\gamma_x$ = 10.0 $ps^{-1}$. For SAMD simulations, $T$ = 100 K was used for the Nosé-Hoover thermostat, with $\delta t$ = 0.5 fs and $N_{steps}$ = 2 × $10^6$. Other parameters were set as $r_D$ = 3.46 Å (first valley in RDF of ideal FCC Al), κ = 0.05 eV/$Å^2$, and $\gamma_s$ = 5.0 $ps^{-1}$. Six SAMD simulations were conducted with $T_s$ = 100, 200, 300, 400, 500, and 600 K, respectively. Additionally, a hybrid MD & Monte Carlo (MC) simulation was performed at 100 K, combining conventional MD with MC random moves of H atoms (1,000 attempts every 1,000 MD steps). No barostat was used in the simulations.

Figs. 6(a) and 6(b) show the time evolution of the instantaneous kinetic temperature for conventional MD ($T$ = 100 K) and SAMD ($T$ = 100 K and $T_s$ = 500 K) simulations, respectively. Profiles for SAMD simulations with other $T_s$ values were very similar to that in Fig. 6(b). The standard deviation (SD) of the instantaneous kinetic temperature, shown in Figs. 6(c) and 6(d), confirms that SAMD simulations maintain excellent temperature control, comparable to conventional MD.

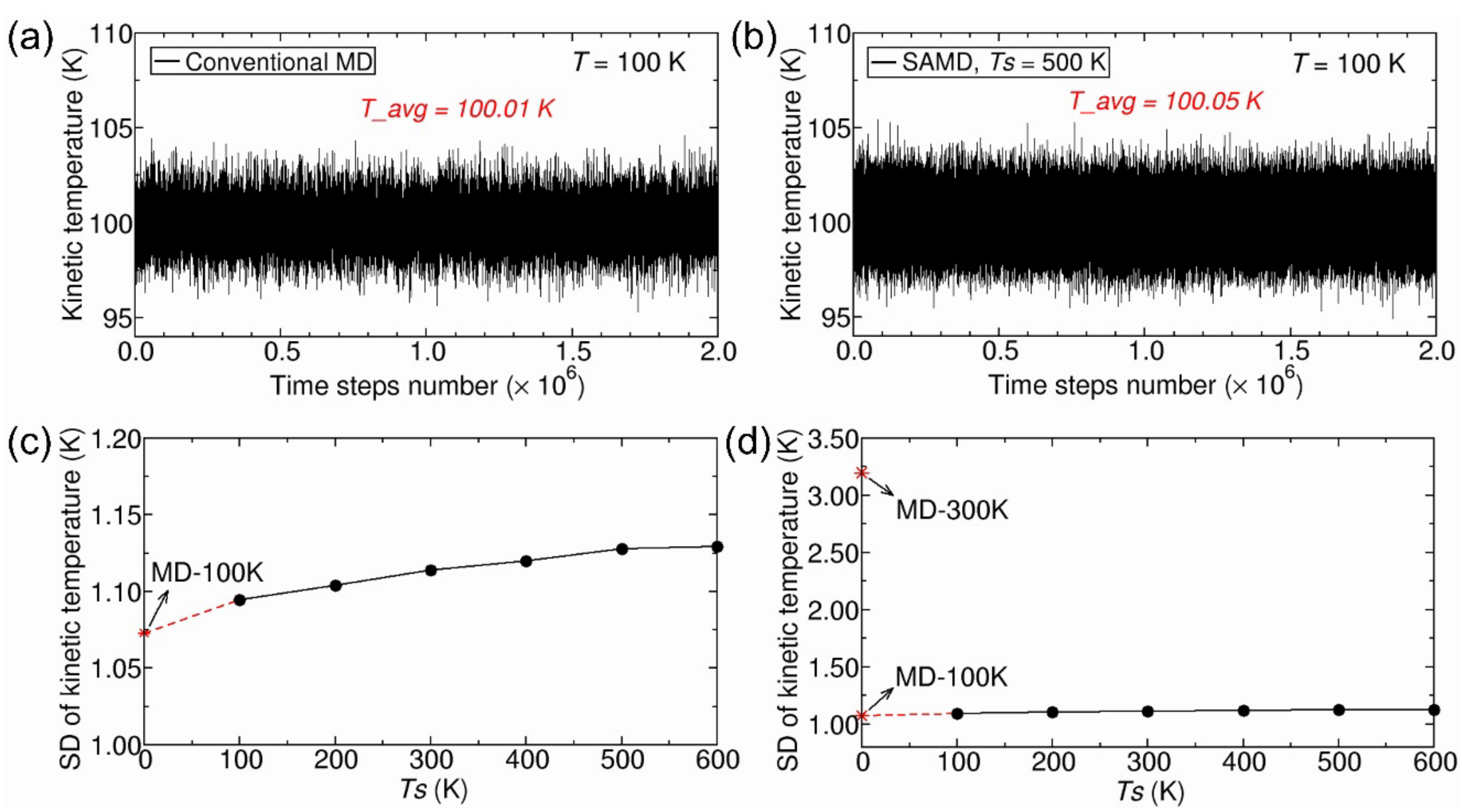

**Fig. 6.** Instantaneous kinetic temperature (Eq. (3)) evolution in the bicrystal model (Fig. 5): (a) Conventional MD at $T$ = 100 K; (b) SAMD at $T$ = 100 K and $T_s$ = 500 K, with time-averaged values (‘$T_avg$’) shown. Data sampled every five time steps. (c) Standard deviation (SD) of instantaneous kinetic temperature over time for SAMD ($T$ = 100 K). (d) SD comparison between conventional MD ($T$ = 100 K or 300 K) and SAMD ($T$ = 100 K).

The RMSD of H atoms from conventional MD and SAMD simulations is shown in Fig. 7. Conventional MD at 100 K shows negligible H diffusion over 10 ns, while significant diffusion occurs at 300 K (Fig. 7(a)). In contrast, SAMD simulations at $T$ = 100 K exhibit steady-state H diffusion with substantial displacement (Fig. 7(b)), demonstrating strong acceleration effect of the SAMD method.

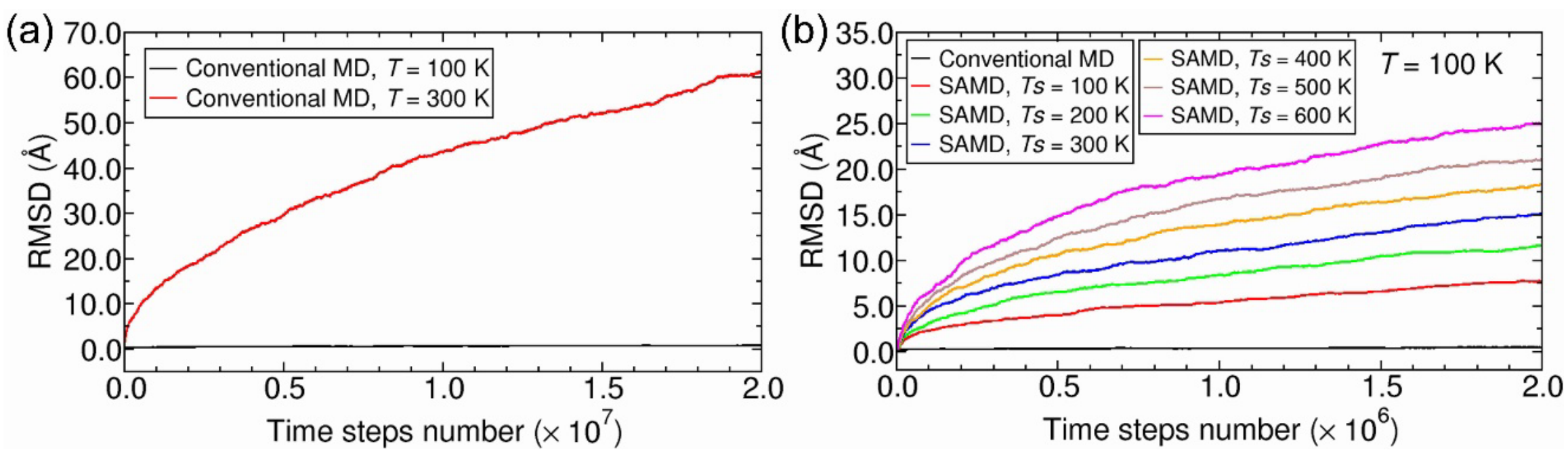

**Fig. 7.** RMSD of H atoms in the bicrystal model (Fig. 5): (a) Conventional MD at $T$ = 100 K and 300 K; (b) SAMD at $T$ = 100 K. Note the difference in time steps ($N_{steps}$) between (a) and (b).

Fig. 8 illustrates the final H atom distributions. Conventional MD at 100 K (Fig. 8(a)) shows almost no change from the initial state, while H segregation at GBs is observed in conventional MD at 300 K (Fig. 8(b)) and hybrid MD & MC at 100 K (Fig. 8(c)). SAMD simulations at $T$ = 100 K (Figs. 8(d)-(i)) yield significant GB segregation, with the degree of segregation increasing with $T_S$, consistent with the diffusion distances in Fig. 7(b). Supplementary Movie S2 further demonstrates H diffusion and segregation dynamics.

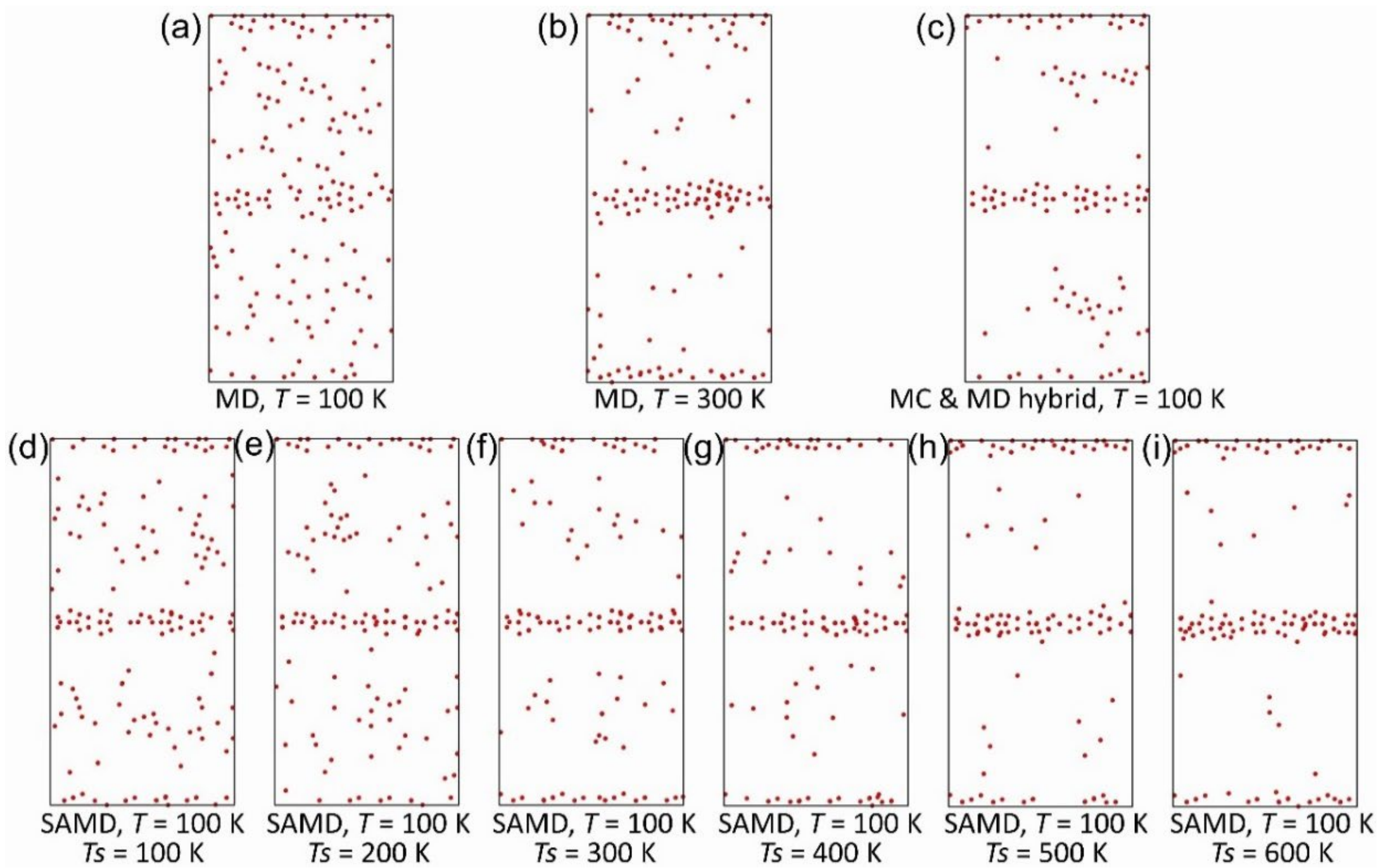

**Fig. 8.** Configurational states of the bicrystal model (Fig. 5) at simulation end: (a, b) Conventional MD; (c) Hybrid MD/MC; (d-i) SAMD. Only H atoms are shown, as structural configurations of Al atoms remained unchanged in all simulations.

H diffusion in the bicrystal involves multiple migration paths with varying energy barriers, leading to non-trivial kinetic behaviors. The agreement in H segregation between SAMD ($T$ = 100 K), conventional MD ($T$ = 300 K), and hybrid MD & MC ($T$ = 100 K) simulations validates the SAMD method's ability to consistently accelerate simulations of H diffusion in Al bicrystals. The acceleration effect can be tuned by adjusting $T_S$. Using Eq. (5), the speedup ratios for SAMD simulations with $T_S$ = 100 K, 200 K, 300 K, 400 K, 500 K, and 600 K are calculated as 918, 2121, 3539, 5216, 6855, and 9691, respectively.

### 4.2. Case study (2): Shear response of a symmetric tilt GB in Al

Fig. 9(a) illustrates an Al bicrystal model with the Σ11<1 $\bar{1}$ 0>{1 3 1} symmetric tilt GB. The model dimensions are 9.9 nm × 8.0 nm × 18.9 nm, containing 89,182 Al atoms. An EAM potential [62] was used to describe Al-Al interactions. Periodic boundary conditions were applied along the X- and Y-directions, while two border slabs at the top and bottom in the Z-direction were designed for shear loading. The optimized GB configuration, obtained through in-plane relative shifts and energy minimization, is shown in Fig. 9(b).

Shear responses along two orthogonal directions parallel to the GB plane ('Shear-XZ' in Fig. 9(c) and 'Shear-YZ' in Fig. 9(d)) were simulated using conventional MD and SAMD methods at $T$ = 300 K. A similar study using only conventional MD was reported previously [63].

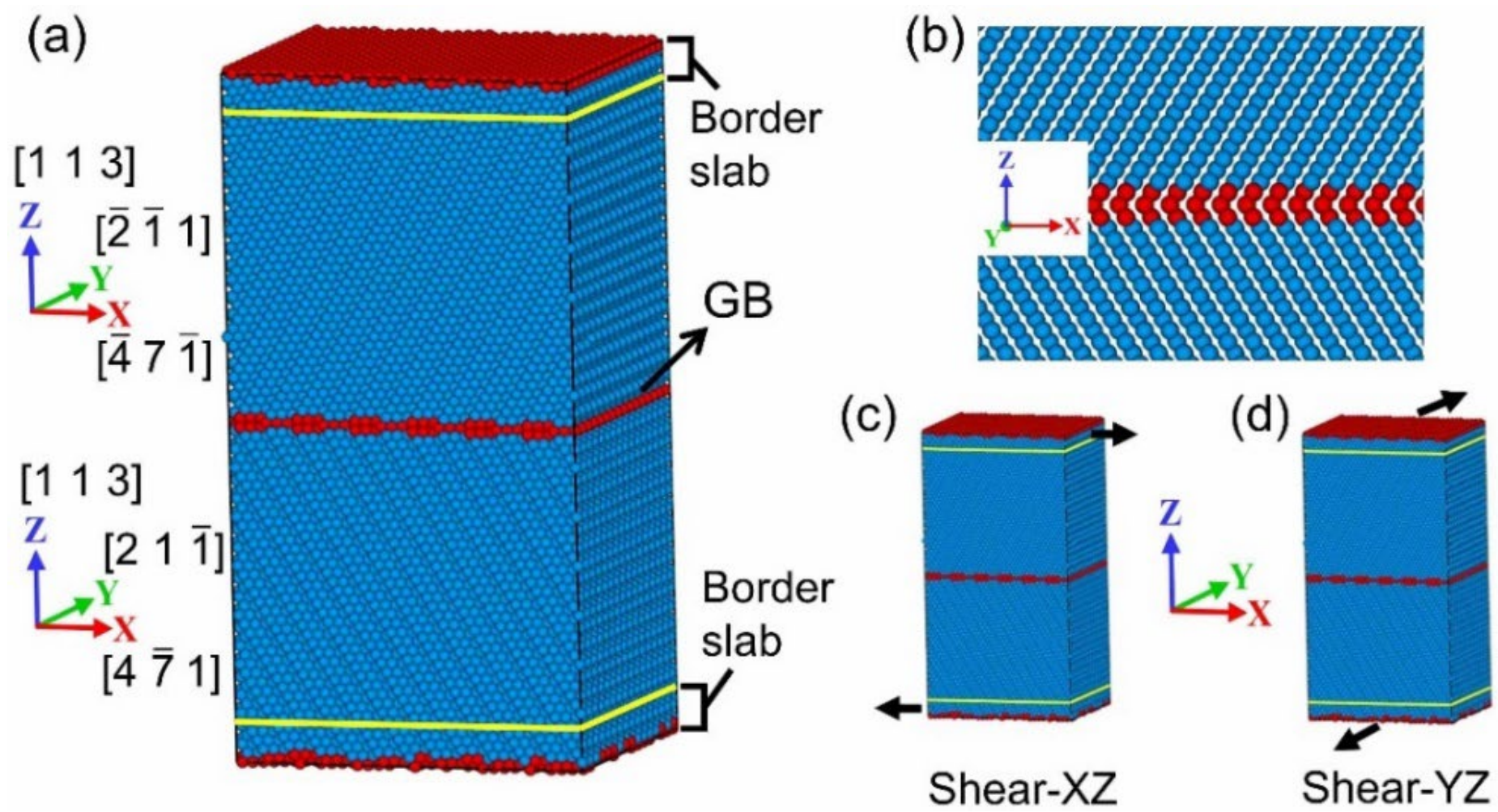

**Fig. 9.** (a) A bicrystal model of the Σ11<1 $\bar{1}$ 0>{1 3 1} symmetric tilt GB in Al, with lattice orientations of the upper and lower crystals indicated. (b) Optimized GB structure viewed along the Y- axis. (c) and (d) Schematic illustrations of shear loading along X- direction ('Shear-XZ') and Y- direction ('Shear-YZ'), respectively. Blue spheres represent Al atoms with FCC local structure, while red spheres represent non-FCC Al atoms, identified using the CNA method.

Conventional MD simulations used a Nosé-Hoover thermostat ($\gamma_x$ = 10.0 ps$^{-1}$) to control the temperature of the inner model at $T$ = 300 K and a Nosé-Hoover barostat [64] (damping coefficient: 2.0 ps$^{-1}$) to maintain zero stress ($\sigma_{XX} = \sigma_{YY} = \sigma_{ZZ}$ = 0 Pa). The atoms in border slabs followed Langevin dynamics at $T$ = 300 K. A time step of $\delta t$ = 1.0 fs and a strain rate of $1 \times 10^8$ s$^{-1}$ were applied, resulting in a final shear strain of 0.2 over $2 \times 10^6$ steps. Shear loading was applied stepwise by shear transformation of the model along X- or Y-axis, with the center-of-mass of atoms in each border slab constrained according to the shear strain. The border slabs were thus allowed in-plane side displacements perpendicular to the shear direction to accommodate eigenstrains from structural transformations, enabling analysis of the transformations through the monitoring of these side displacements [63].

For SAMD simulations, the same barostat and shear loading were applied. Six SAMD simulations with $T_s$ = 500, 1000, 1500, 2000, 4000, and 6000 K were performed for each shear direction. Other parameters were $r_D$ = 3.46 Å (first valley in RDF of ideal FCC Al), κ = 0.05 eV/Å$^2$, $T$ = 300 K, $\gamma_x$ = 100.0 ps$^{-1}$, and $\gamma_s$ = 5.0 ps$^{-1}$. A time step of $\delta t$ = 0.5 fs and a constant strain rate were applied to give the same final strain of 0.2 over 4 × 10$^6$ steps, corresponding to an apparent strain rate of 1 × 10$^8$ s$^{-1}$ without considering the acceleration effect.

The SAMD method maintained excellent temperature control during shear loading, with the instantaneous kinetic temperature closely matching the target value of $T$ = 300 K for all $T_s$. The stress-strain curves (Figs. 10(a) and 10(c)) exhibit sawtooth-like stress relaxation behavior, due to the nucleation and glide of GB displacement shift complete (DSC) dislocations (Figs. 11(a) and 11(b)). These dislocations are characterized by the Burgers' vector and step height, and can induce slide-migration coupled motion of GB, as illustrated in Figs. 11(c) and 11(d) and Supplementary Movies S3a and S3b. Finite in-plane side displacements of the border slabs occurred during shear along the Y-direction (Fig. 10(d)) but not the X-direction (Fig. 10(b)), due to the alignment of the shear direction with the Burgers' vector of the nucleated GB DSC dislocations. These findings align with previous conventional MD studies [63].

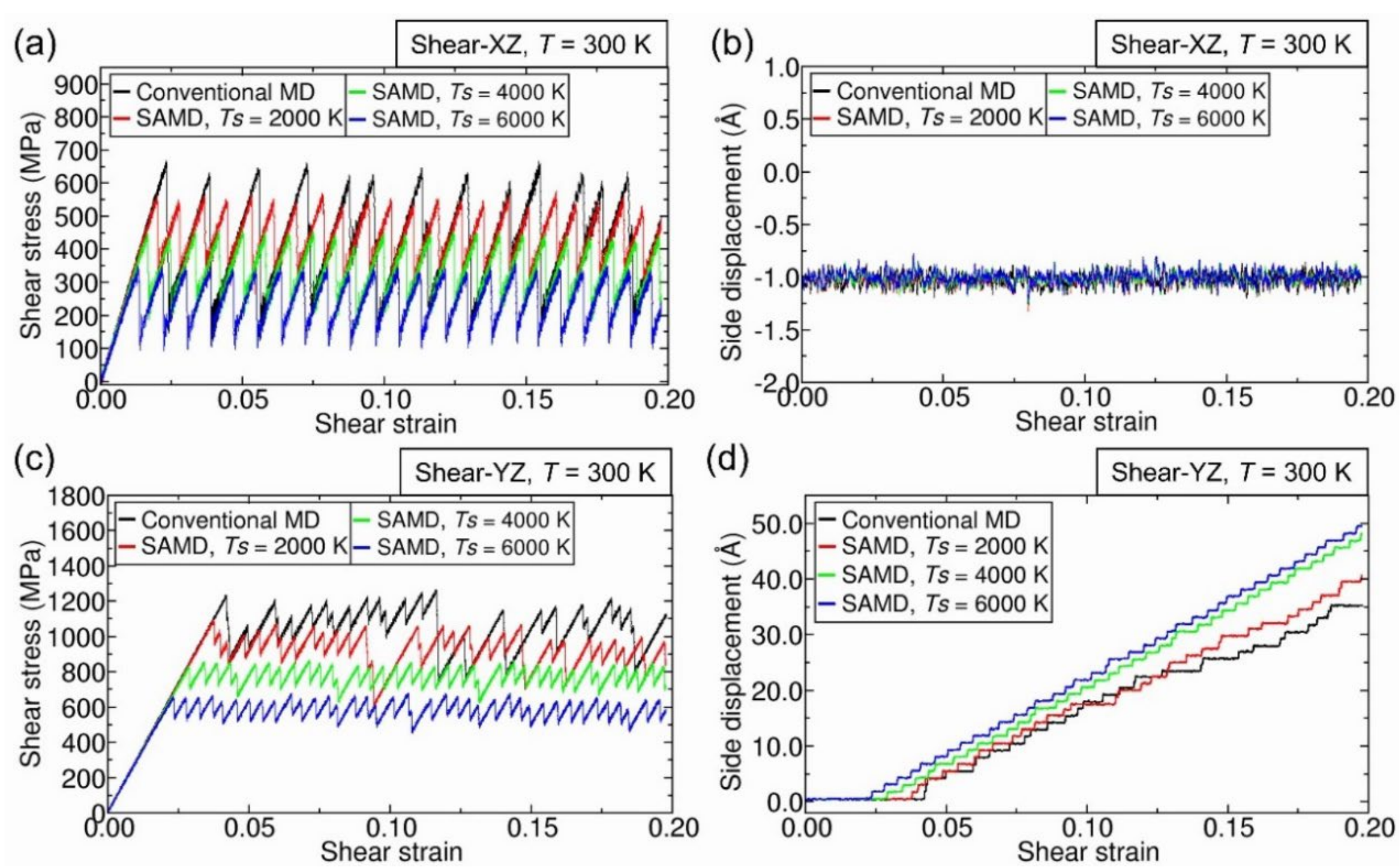

**Fig. 10.** (a, c) Stress-strain curves for shear of the bicrystal model (Fig. 9) along the X- and Y-direction, respectively. (b, d) In-plane side displacement of border slabs versus strain for shear

along the X- and Y-direction, respectively. Note that only the results with $T_S$ = 2000, 4000, and 6000 K for SAMD simulations are displayed in the figure for clarity.

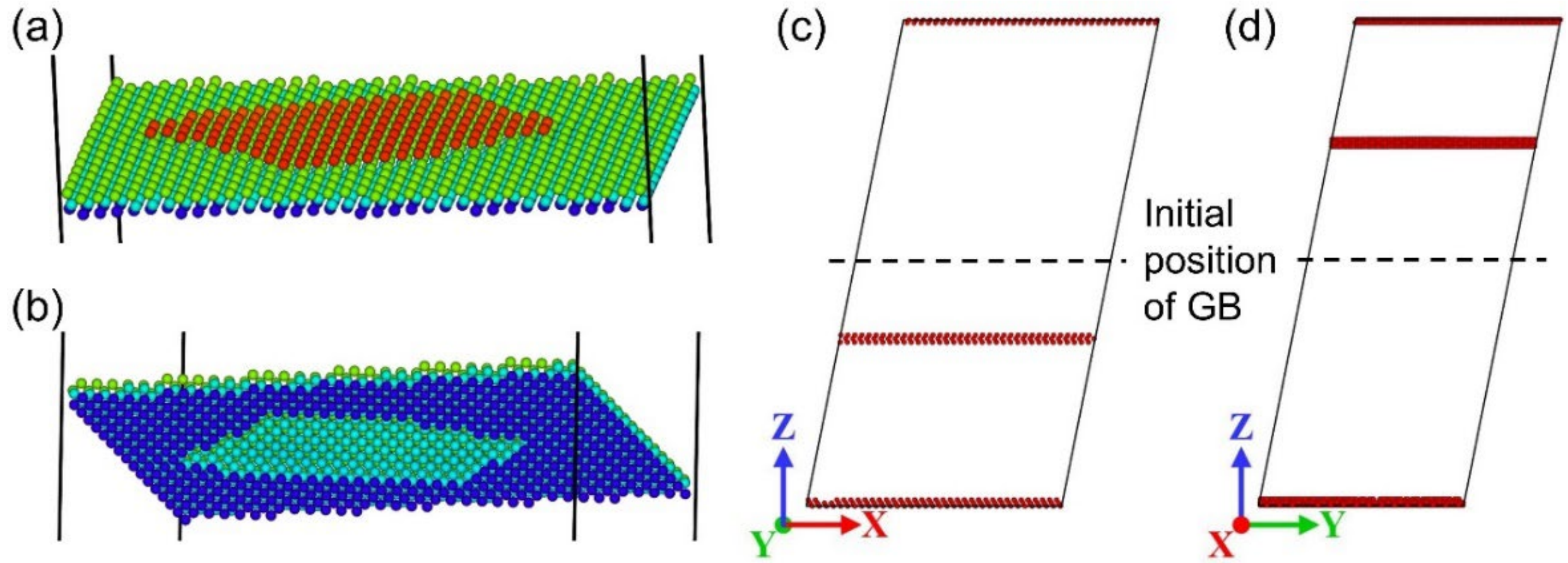

**Fig. 11.** (a, b) Top and bottom views of a GB DSC dislocation nucleated and expanding through the GB plane during Y-direction shear, with Burgers' vector of (1/22) [4 $\bar{7}$ 1] and step height equal to the (1 1 3) interplane distance of Al. (c, d) Final GB state after SAMD simulations ($T_S$ = 6000 K) for shear along the X- and Y-directions, respectively. Only non-FCC structured atoms are shown. In (a, b), red, green, cyan, and blue spheres indicate atoms at progressively lower heights along the Z-axis.

The peak stresses in the stress-strain curves (Figs. 10(a) and 10(c)) correspond to the critical shear stresses for GB DSC dislocation nucleation, scaled by Schmid factors [63]. Conventional MD simulations, constrained by their inherent timescale limitations, employed an extremely high strain rate ($1 \times 10^8$ s$^{-1}$), yielding peak stresses of ~660 MPa (X-direction) and ~1200 MPa (Y-direction). In contrast, SAMD simulations monotonically reduced these peak stresses with increasing $T_S$, reflecting a significant timescale extension and reduction in the effective strain rate at nearly the same computational cost, while preserving the underlying structural transition mechanisms.

The speedup ratios of SAMD simulations can be quantified by comparing the effective strain rates and simulation time steps per strain with those of conventional MD. To estimate the effective strain rate, we employed a statistical mechanics model for surface dislocation nucleation [65]. Assuming a linear stress dependence of activation energy for GB DSC dislocation nucleation, an approximate relationship between the shear strain rate $\dot{\varepsilon}$ and the peak shear stress $\tau$ can be expressed as:

$$\ln \dot{\varepsilon} = A\tau + B \ , \tag{6}$$

where $A$ and $B$ are constants that depend on the GB DSC dislocation type (characterized by the Burgers' vector and step height) and temperature. We then conducted conventional MD simulations for both shear directions at four strain rates of $1 \times 10^9$, $1 \times 10^8$, $1 \times 10^7$, and $1 \times 10^6$ $s^{-1}$ to fit $A$ and $B$ (Table 2). For shear in the X-direction, the fitted values are $A = 0.201$ and $B = -112.5$; for shear in the Y-direction, the values are $A = 0.068$ and $B = -67.3$.

**Table 2**

Averaged peak stresses on stress-strain curves for shear of the model in Fig. 9(a) along the X- and Y-directions at 300 K, obtained from conventional MD simulations. For each strain rate, five simulations with different random seeds in Langevin dynamics for border slab atoms were performed. Only the first peak stress on each curve was used for averaging.

| Strain rate ($s^{-1}$) | Peak stress (MPa) (Shear-XZ) | Peak stress (MPa) (Shear-YZ) |
|---|---|---|
| $1 \times 10^9$ | 665.54 ± 18.14 | 1301.97 ± 8.91 |
| $1 \times 10^8$ | 650.65 ± 3.24 | 1260.69 ± 23.18 |
| $1 \times 10^7$ | 640.80 ± 3.62 | 1228.15 ± 12.62 |
| $1 \times 10^6$ | 631.01 ± 2.59 | 1200.51 ± 11.83 |

The effective strain rates were estimated by applying the peak stresses on the stress-strain curves derived from SAMD simulations to Eq. (6), as summarized in Table 3. Within the approximation error of Eq. (6), the calculated strain rates for both shear directions show good consistency for $T_s$ = 500, 1000, 1500 and 2000 K, demonstrating uniform acceleration of the SAMD method with $T_s \leqslant 2000$ K. However, for $T_s$ = 4000 and 6000 K, the effective strain rates diverged significantly, suggesting these values exceed the upper limit for $T_s$ in SAMD simulations of mechanical loading in this system. Notably, the SAMD method achieves an acceleration of ~8 orders of magnitude at $T_s$ = 2000 K.

**Table 3**

The effective strain rates (in unit of $s^{-1}$) of SAMD simulations calculated for both shear directions at $T_s$ = 500, 1000, 1500, 2000, 4000 and 6000 K, respectively.

| Shear direction | $T_s$= 500 K | $T_s$= 1000 K | $T_s$= 1500 K | $T_s$= 2000 K | $T_s$= 4000 K | $T_s$= 6000 K |
|---|---|---|---|---|---|---|
| X- | $3.6 \times 10^5$ | $1.4 \times 10^3$ | 31 | 0.04 | $2.0 \times 10^{-11}$ | $2.5 \times 10^{-20}$ |
| Y- | $3.3 \times 10^7$ | $1.6 \times 10^5$ | $1.9 \times 10^3$ | 23 | $3.4 \times 10^{-5}$ | $7.3 \times 10^{-11}$ |

### 4.3. Case study (3): Tensile behavior of α-Fe bicrystal

Figs. 12(a) and 12(b) show a bicrystal model of the Σ9 <1 $\bar{1}$ 0> pure twist GB in α-Fe, with dimensions 10.3 nm × 9.7 nm × 20.2 nm and 172,800 Fe atoms. An EAM potential [66,67] was used to describe Fe-Fe interactions. Periodic boundary conditions in all three dimensions result in two identical GBs ('GB-1' and 'GB-2'). The optimized GB configuration, obtained using the same exploration procedure as in Case Study (2), is shown in Fig. 12(c). Tensile loading along the Z-direction at $T$ = 300 K was simulated using conventional MD and SAMD methods. A similar study using only conventional MD was reported previously [59].

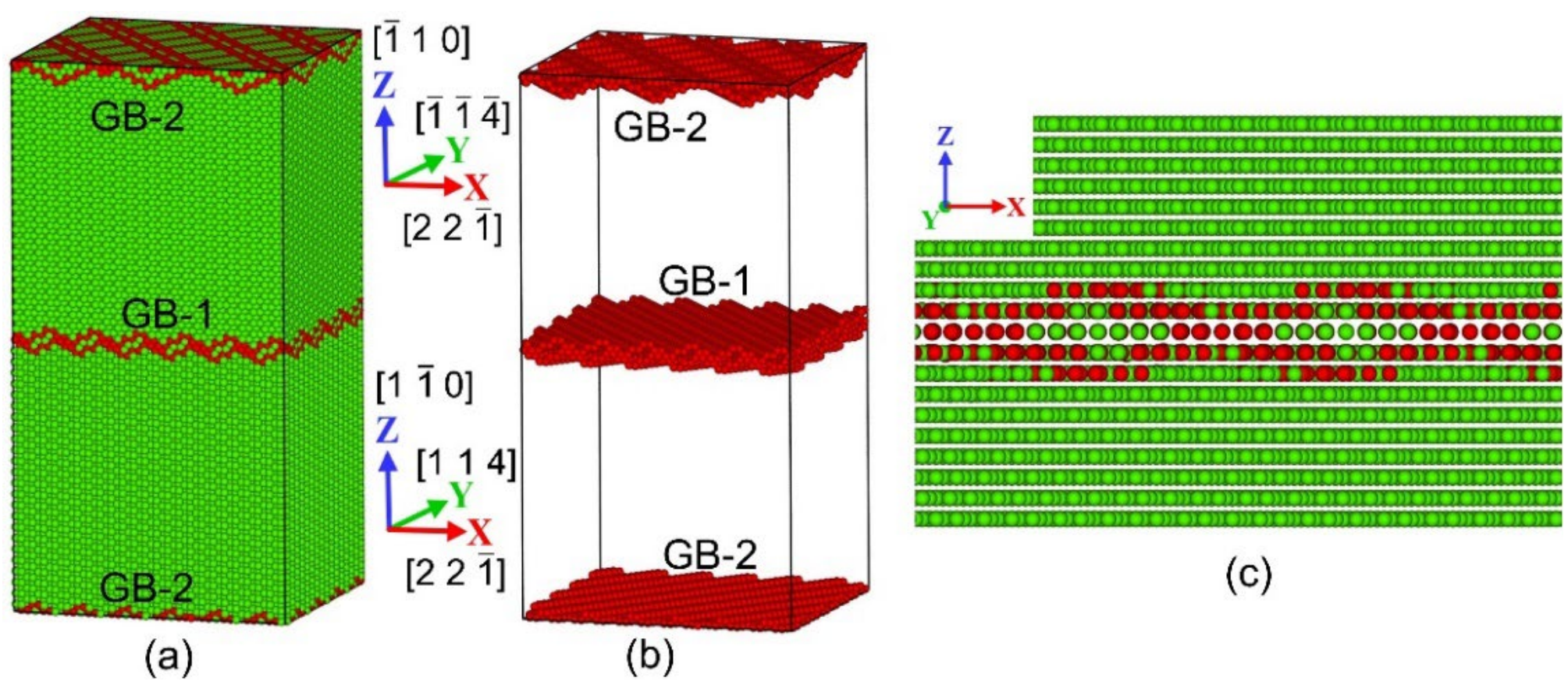

**Fig. 12.** (a, b) A bicrystal model of the Σ9 <1 $\bar{1}$ 0> pure twist GB in α-Fe, with lattice orientations of the upper and lower crystals indicated. Periodic boundary conditions in all three dimensions result in two identical GBs ('GB-1' & 'GB-2'). (c) Optimized GB structure viewed along the Y-axis. Green and red spheres represent atoms of BCC and non-BCC local structure, respectively. In (b), BCC atoms are omitted.

Conventional MD simulations employed a Nosé-Hoover thermostat ($\gamma_x$ = 5.0 ps$^{-1}$) to maintain $T$ = 300 K. A time step of $\delta t$ = 2.0 fs and a strain rate of 1 × $10^7$ s$^{-1}$ yielded a final tensile strain of 0.4 over $2 \times 10^7$ steps. Tensile loading was applied stepwise along the Z-axis, with a Nosé-Hoover barostat [64] (damping coefficient: 2.0 $ps^{-1}$) maintaining $\sigma_{XX} = \sigma_{YY} = 0$ Pa, simulating uniaxial tensile conditions.

For SAMD simulations, the same loading protocol and barostat were applied. Three SAMD simulations with $T_s$ = 3000, 5000, and 7000 K were performed, with other parameters set as $r_D$ = 3.46 Å (first valley in RDF of ideal BCC Fe), κ = 0.05 eV/Å$^2$, $T$ = 300 K, $\gamma_x$ = 100.0 $ps^{-1}$, and $\gamma_s$ = 5.0 $ps^{-1}$. A time step of $\delta t$ = 0.5 fs and $N_{steps}$ = $8 \times 10^6$ were used to achieve the same final strain of 0.4, corresponding to an apparent strain rate of $1 \times 10^8$ $s^{-1}$ without considering the acceleration effect.

Excellent temperature control was maintained in all SAMD simulations. The stress-strain curves (Fig. 13) exhibit serrated profiles due to stress relaxations triggered by various structural transitions during tensile loading. These transitions include dislocation nucleation from GBs, dislocation-GB interactions (impingement, absorption, or transmission), and dislocation reactions, as shown in Fig. 14 and Supplementary Movie S4. Similar results were reported in previous conventional MD studies [59].

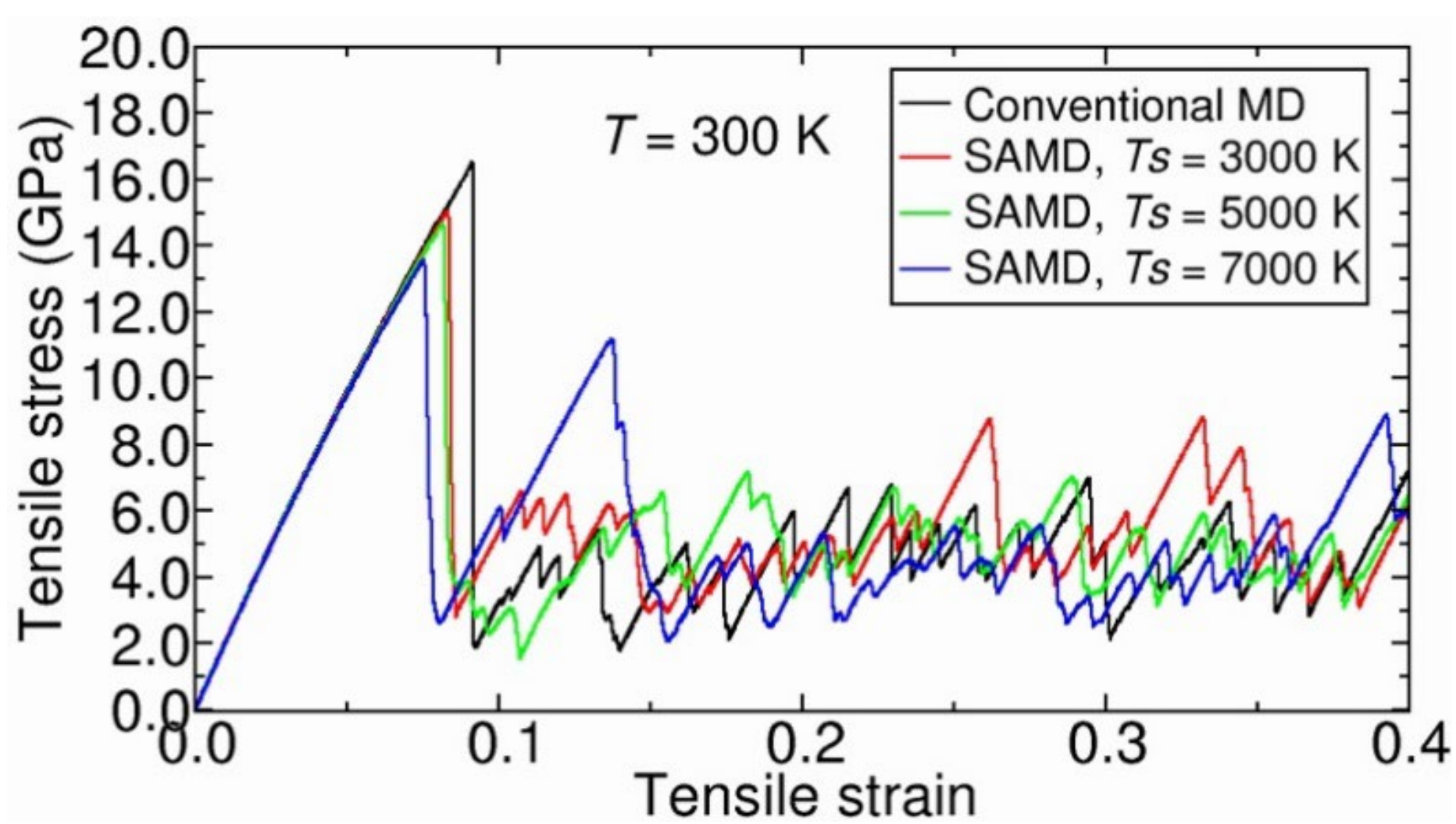

**Fig. 13.** Stress-strain curves for uniaxial tensile loading of the bicrystal model (Fig. 12) along the Z-direction at 300 K, obtained from conventional MD and SAMD simulations with varying $T_s$.

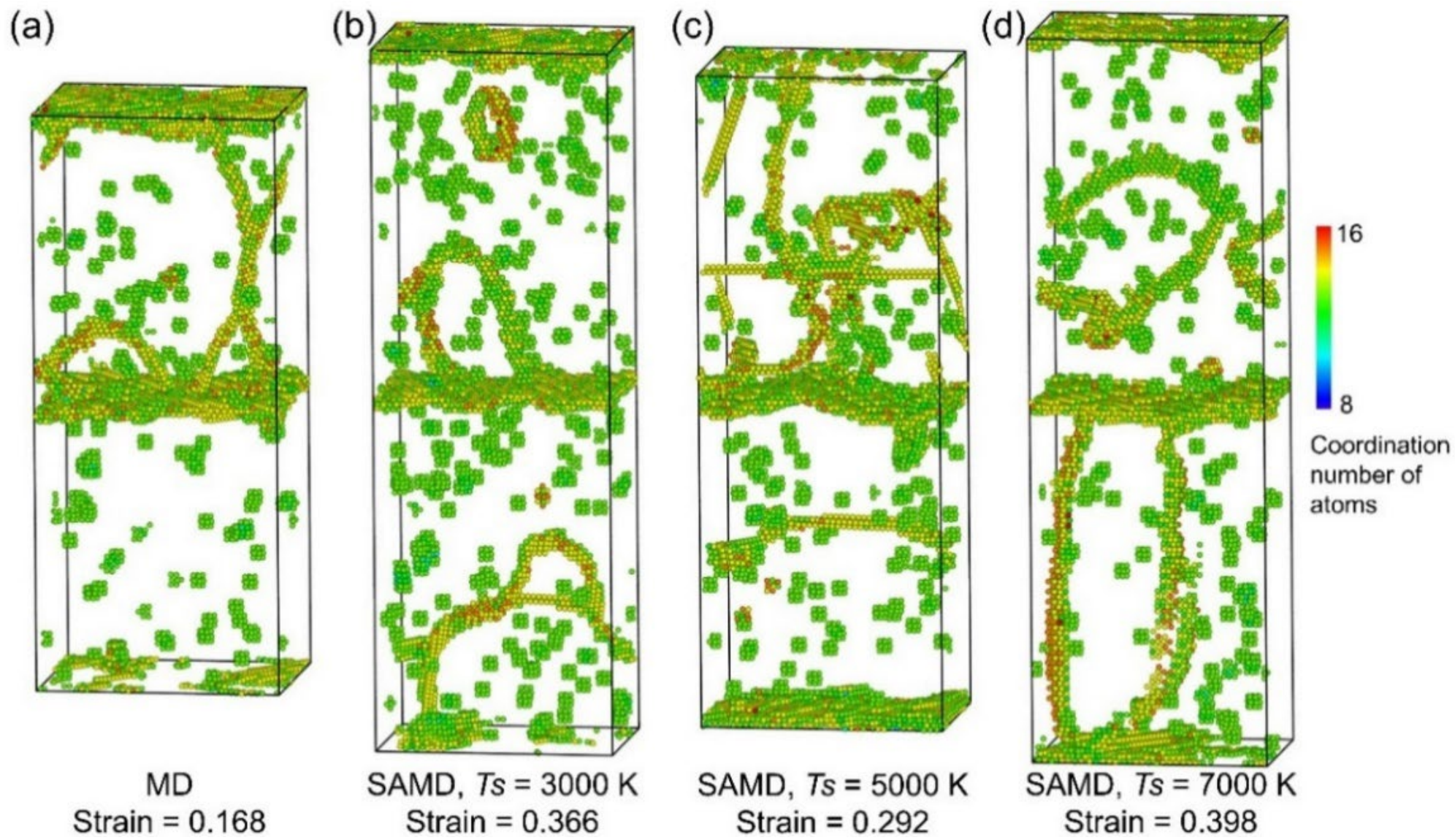

**Fig. 14.** Defect structures in the bicrystal model during tensile loading at $T$ = 300 K and varying strain levels: (a) Conventional MD; (b-d) SAMD simulations with $T_s$ = 3000, 5000, and 7000 K. Only non-BCC atoms are shown, colored by coordination number (cutoff radius: 3.46 Å, see color bar). The corresponding tensile strain for each image is indicated.

In conventional MD simulations, the first peak stress of 16.5 GPa on stress-strain curve corresponds to dislocation nucleation from a GB. This thermally activated process is strain-rate dependent, with higher strain rates requiring larger critical stresses. The extremely high strain rate ($1 \times 10^7$ s$^{-1}$) in conventional MD simulation, necessitated by its timescale limitations, leads to this high peak stress. SAMD simulations significantly reduce the first peak stress with increasing $T_s$, demonstrating effective acceleration and timescale extension for achieving much lower strain rates. A comparison of stress-strain curves (Fig. 13) and defect structures (Fig. 14, Supplementary Movie S4) between conventional MD and SAMD simulations reveals similar characteristics of structural transitions, confirming consistent acceleration in tensile loading simulations of the α-Fe bicrystal using the SAMD method.

### 4.4. Case study (4): Surface diffusion of Ag atoms on Ag film

To evaluate the method's applicability in studying surface microstructure evolution, we designed an Ag thin film model (Fig. 15). Periodic boundary conditions were applied along the X- and Y-directions, with a rigid border slab at the bottom simulating the thin film.

Three surface point defects — a mono vacancy, an Ag adatom, and an Ag add-on dimer — were introduced to study surface diffusion. The model dimensions (Lx × Ly × Lz) are 6.1 nm × 8.2 nm × 3.4 nm, containing 9,602 Ag atoms. An EAM potential [68] was adopted to describe the Ag-Ag interactions. The presence of these point defects enables multiple structural transition paths with varying energy barriers and kinetic characteristics for surface diffusion. Atom coloring in Fig. 15 aids in tracking and distinguishing migration paths.

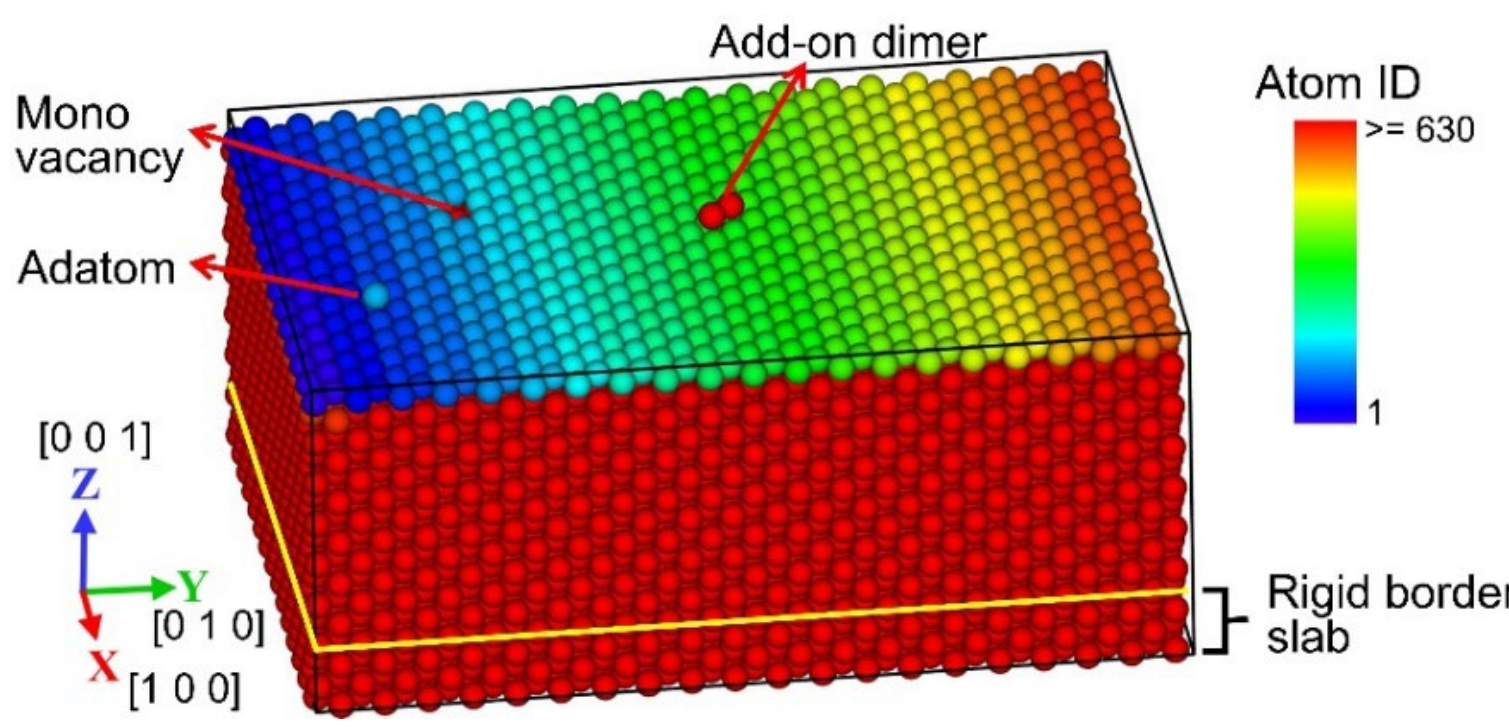

**Fig. 15.** Atomistic model of a Ag thin film, featuring a mono vacancy, a Ag adatom and a Ag add-on dimer on the surface. The bottom five atom layers are rigid. Surface atoms (IDs 1~630) are colored by ID, while atoms in the add-on dimer and subsurface layers (IDs > 630) are uniformly red.

Conventional MD simulations of surface diffusion were performed at four temperatures ($T$ = 325, 500, 700, and 800 K) using a Nosé-Hoover thermostat ($\gamma_x$ = 5.0 ps$^{-1}$). A time step of $\delta t$ = 1.0 fs and $N_{steps}$ = 5 × $10^6$ were used, corresponding to 5 ns diffusion process. No barostat was applied. For SAMD simulations, $T$ = 325 K was used, with $T_s$ = 3000, 4000, 5000, and 6000 K. Other parameters were $r_D$ = 3.49 Å (first valley in RDF of ideal FCC Ag), κ = 0.05 eV/Å$^2$, $\gamma_x$ = 100.0 ps$^{-1}$, and $\gamma_s$ = 5.0 ps$^{-1}$. A time step of $\delta t$ = 0.5 fs and $N_{steps}$ = 1 × $10^7$ were used, with no barostat as well.

Again, SAMD simulations maintained excellent control of the temperature. The RMSD curves for Ag atom diffusion (Fig. 16) show that the overall migration distance increases with temperature in conventional MD simulations and with $T_s$ in SAMD simulations at $T$ = 325 K. The limited number of surface defects and their interactions (e.g., vacancy-adatom annihilation) introduce irregularities in the RMSD curves.

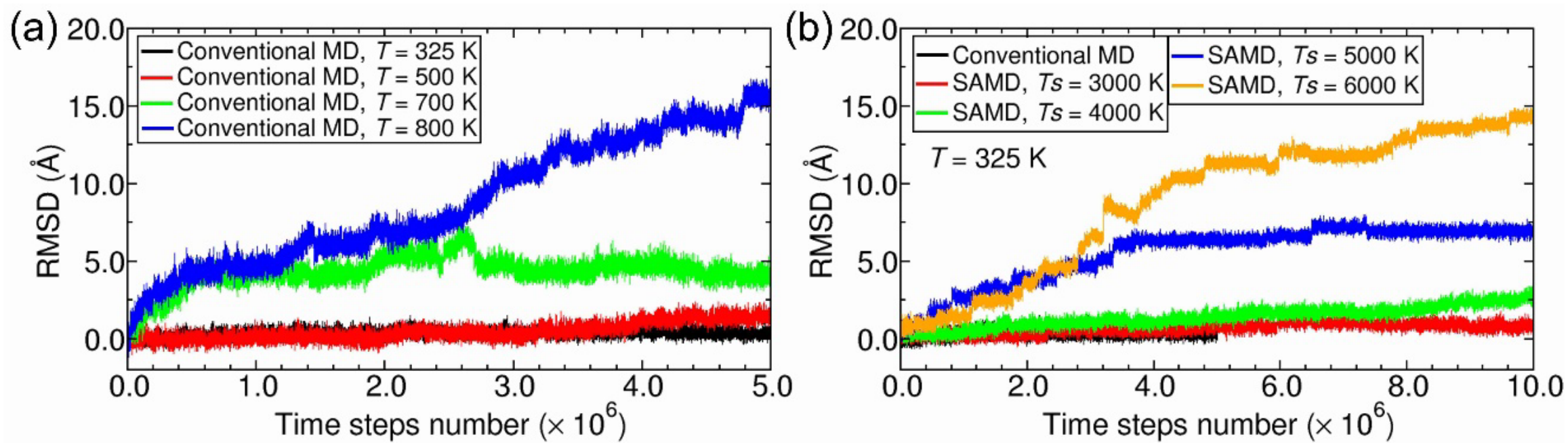

**Fig. 16.** RMSD of Ag atoms in conventional MD (a) and SAMD (b) simulations of the Ag thin film model (Fig. 15).

As shown in Figs. 17 and 18, Ag atom diffusion occurs through point defect migration on the film surface in both conventional MD and SAMD simulations. Supplementary Movies S5a and S5b (see Appendix A) illustrate the time evolution of surface structure and point defect transformations. Notably, adatom and dimer migration can follow multiple paths with varying energy barriers, such as collective or individual atom hops. Additionally, point defect transformations, including dimer separation into adatoms and vacancy-adatom annihilation, were observed.

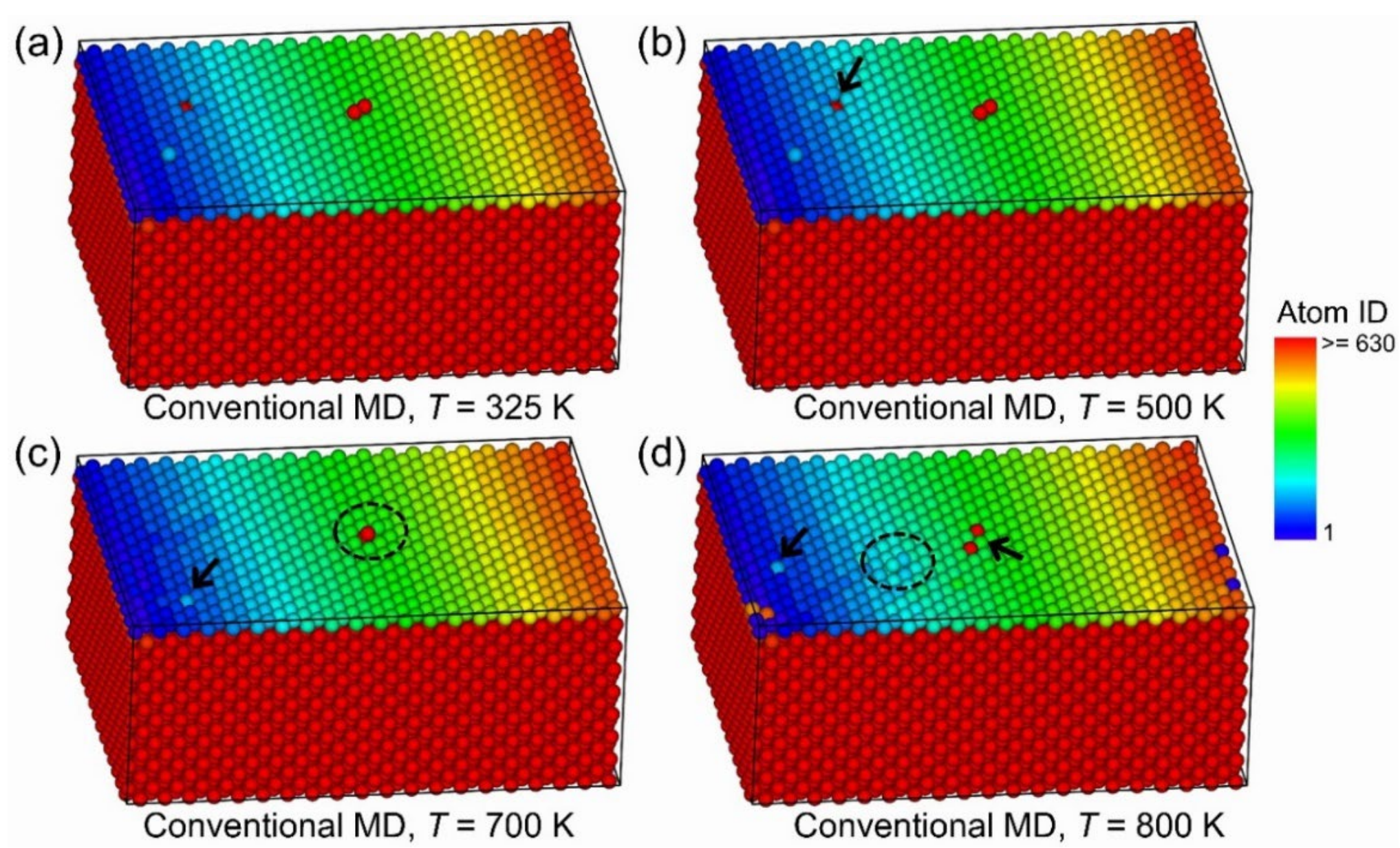

**Fig. 17.** Configurational states of the Ag thin film after conventional MD simulations at $T$ = 325 K (a), 500 K (b), 700 K (c), and 800 K (d). Sphere coloring follows Fig. 15. The circles in (c) and (d) mark the surface dimer. The arrows indicate traces of point defect migration or transformation.

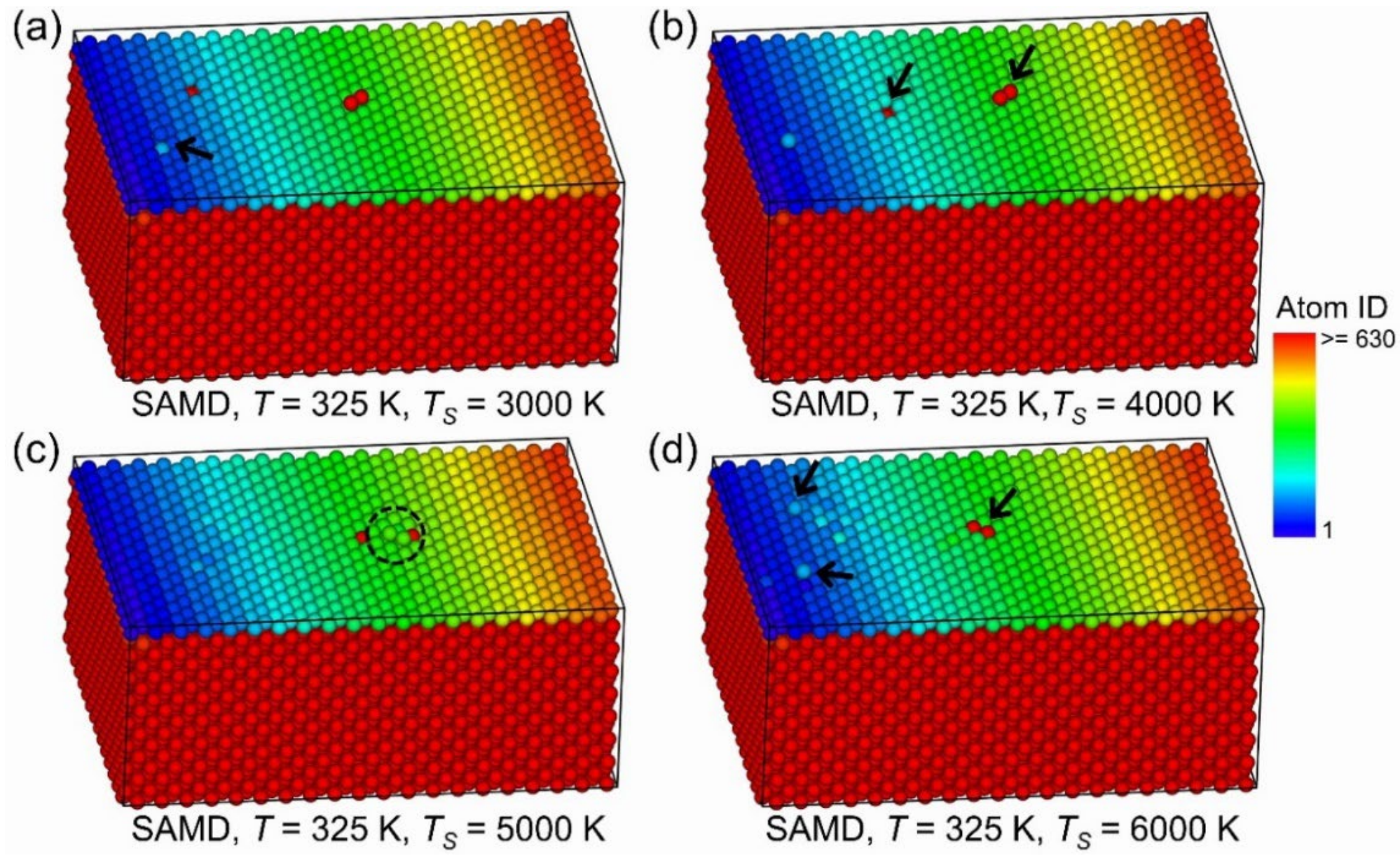

**Fig. 18.** Configurational states of the Ag thin film after SAMD simulations at $T$ = 325 K for $T_S$ = 3000 K (a), 4000 K (b), 5000 K (c), 6000 K (d). Sphere coloring follows Fig. 15. The circle in (c) marks a surface dimer. The arrows indicate traces of point defect migration or transformation.

The significantly increased migration distance in SAMD simulation at $T$ = 325 K compared to conventional MD at the same temperature (Fig. 16(b)) highlights the acceleration effect of the SAMD method. Furthermore, the structural transitions observed in SAMD simulations resemble those in conventional MD at higher temperatures (Figs. 17 and 18, Supplementary Movies S5a and S5b). Although temperature dependence of rate of different structural transitions can vary, we argue that this resemblance indicates consistent acceleration in SAMD simulations of Ag surface diffusion.

## 5. Discussion

### 5.1. Perspective from current simulation practices using the SAMD method

The SAMD method employs NNOADs of all atoms, as defined in Eq. (1), as a generalized reaction coordinate for various kinds of structural transitions in materials, along with extended system dynamics outlined in Eq. (2) to describe atomic motions. This extended system dynamics integrates Nosé-Hoover dynamics to thermostat the system and Langevin dynamics for evolution of three additional dynamic variables of each atom, harmonically coupled to the three NNOAD components of the atom. The method has only a few well-defined parameters, whose proper values can be determined through preliminary dynamic analysis and benchmark simulations. Specifically, the migration of two types of point defects in α-Fe crystals serves as an effective benchmark, where acceleration effects quantified by Eq. (5) enable parameter optimization through consistency criteria.

By applications to four typical materials science problems — H segregation at GB in Al, shear response of GB in Al, tensile loading of α-Fe bicrystal, and Ag surface diffusion — alongside the benchmark point defects diffusion study, SAMD demonstrates significant acceleration of microstructure evolution compared to conventional MD while maintaining kinetic consistency of structural transitions. The degree of acceleration, controlled by parameter $T_s$ in Eq. (2), can reach up to 8 orders of magnitude in certain cases. Importantly, SAMD simulations consistently yield physically reasonable microstructure evolution behaviors, validated against conventional MD results. This confirms the ability of SAMD to accurately predict both the occurrence and temporal sequence of structural transitions under given thermodynamic conditions, achieving kinetic consistency in all cases studied.

These findings suggest the empirical formulation of SAMD holds substantial promise for accelerated MD simulations of a broad range of materials science problems. The effectiveness of the method stems from the fact that practical microstructure evolution in materials — whether involving dislocation activities, grain boundary dynamics, precipitations, phase transformations, or other processes — can generally be decomposed into a series of elementary structural transition events like point defect diffusion or local atomic rearrangements. SAMD appears capable of correctly sequencing these elementary structural transition events under usual thermodynamic conditions, as evidenced by our case studies. This capability enables accelerated computer simulations with substantially reduced wall time (potentially by several orders of magnitude) compared to conventional MD, using identical computational resources. Consequently, SAMD opens possibilities for atomistic simulation study of various materials processes with relatively long timescale — such as ordinary deformation of materials, alloy aging, radiation damage evolution, and vapor crystal growth — that remain inaccessible to conventional MD due to inherent severe timescale limitation.

### 5.2. Some preliminary thoughts on theoretical justification of the SAMD method

Nevertheless, several issues remain unresolved. The effectiveness and consistency of acceleration of the SAMD method have only been demonstrated through case studies (including the benchmark problem study). They need rigorous theoretical justification. Below, we outline some preliminary thoughts on such a justification, though these remain much incomplete.

Firstly, the use of NNOADs of all atoms as a generalized reaction coordinate for studying various kinds of microstructure evolution of materials requires further validation. Structural transitions in materials inherently consist of relative displacements between the atoms, and NNOAD of an atom, as defined in Eq. (1) and illustrated in Fig. 1, quantifies the deviation of an atom from the geometric center of its nearest neighbors. This suggests that NNOADs of all atoms can effectively capture relative atomic displacements of any structural transition of materials, provided an appropriate cutoff distance $r_D$ (Eq. (1)) is chosen for identifying nearest neighbors of atoms. While a rigorous proof is currently lacking, we consider this a reasonable approximation, pending further validation through extensive simulation studies. Since its validity hinges on the choice of $r_D$ in Eq. (1), future work should examine its impact on the performance of SAMD. Additionally, alternative formulations that may better serve as a generalized reaction coordinate for microstructure evolution of materials could also be explored.

Secondly, we have pointed out that, the dynamic features of the equations for time evolution of $\boldsymbol{X}$ and $\boldsymbol{S}$ (Eq. (2)), is crucial for proper functioning of the SAMD method. Therefore, we believe a comprehensive analysis of the dynamic characteristics of these equations and the role played by using NNOADs of atoms in these equations, is essential to establish the effectiveness and acceleration consistency of the method.

Drawing from Kramers' reaction rate theory [69,70], the evolution of a many-atom system along a reaction coordinate can be modeled as Brownian motion of a fictitious particle under a force field and frictional damping. Since the NNOADs of all atoms serve as a generalized reaction coordinate and their evolution follows Langevin type dynamics [71], it is reasonable that a Langevin dynamic excitation can be exerted on NNOADs of atoms to drive the system over potential energy barriers. This is achieved by harmonically coupling each NNOAD component to an extra dynamic variable governed by Langevin dynamics. This 'driven effect' introduces extra energy, necessitating a thermostat applied to dissipate the energy introduced. The Nosé-Hoover dynamics, known for its effectiveness as a thermostat method in MD simulations [3, 4], is well-suited for this purpose. Its ability to preserve dynamic properties of a many-atom system in Newtonian dynamics, such as time correlation functions, while maintaining the system at target temperature, makes it an ideal choice. This qualitative reasoning leads to the Eqs. (1) and (2) of the SAMD method.

The Langevin dynamic excitation enhances NNOAD fluctuations of atoms at elevated $T_s$, increasing the apparent frequencies of structural transitions. Meanwhile, the Nosé-Hoover thermostat maintains the system temperature at target value (see Fig. 6), suggesting a statistical orthogonality between NNOAD fluctuations and Cartesian vibrations of the atoms. However, a higher damping coefficient $\gamma_x$ is typically required in Nosé-Hoover dynamics to maintain the system temperature, since there can be alignment between these two types of motions. A further quantitative analysis of the dynamic equations of SAMD may benefit from stochastic dynamics theory [71, 72], which we leave for future work.

Thirdly, the extra dynamic variable $\boldsymbol{S}$ introduced in Eq. (2) is $3N$ dimensional, where $\boldsymbol{S} = (\boldsymbol{s}_1, \boldsymbol{s}_2, \dots \boldsymbol{s}_N)$, $\boldsymbol{s}_i = (s_i^{\mathrm{X}}, s_i^{\mathrm{Y}}, s_i^{\mathrm{Z}})$, and $N$ is the number of atoms in system. The atoms can thus be viewed as residing in a six-dimensional space of X-Y-Z-$\mathrm{S^X}$-$\mathrm{S^Y}$-$\mathrm{S^Z}$. Therefore, Eq. (2) essentially rules that, the atoms in the system evolve in this six-dimensional space under the action of a synthesized force field defined by $U_\kappa(\boldsymbol{X}, \boldsymbol{S})$ in Eq. (2). The atomic motions in the X-Y-Z subspace ('real space') follow deterministic Nosé-Hoover dynamics, while motions in the $\mathrm{S^X}$-$\mathrm{S^Y}$-$\mathrm{S^Z}$ subspace are governed by stochastic Langevin dynamics. In mathematics and physics, higher-dimensional extensions of space often give rise to new structures, emergent phenomena, or deeper insight. String theory, for example, posits more than three spatial dimensions, and is considered as a leading candidate for a fundamental theory of physics [73-75]. Analogously, given the orthogonality between NNOAD fluctuations and Cartesian atomic vibrations, we suggest that extending space from three to six dimensions as proposed in this work, provides a means of folding time in many-atom dynamics, thus enabling accelerated computational MD simulations.

Finally, we expect that with a comprehensive theoretical analysis of the SAMD method to be achieved in future, the proper values of $r_D$, κ, $\gamma_x$, $\gamma_s$ and $T_s$ in Eqs. (1) and (2) can be determined precisely, so is the speedup ratio of SAMD relative to conventional MD in simulations.

## 6. Concluding remarks

In this work, we propose an empirical formulation termed the *Shuffling Accelerated Molecular Dynamics (SAMD)* method, which is adapted from the TAMD approach, to enable efficient atomistic simulations and predictions of microstructure evolution in materials. The SAMD method is based on the approximation that the collection of nearest neighbor off-centering absolute displacements (NNOADs) of all atoms (Eq. (1)), a $3N$-dimensional vector for a system of $N$ atoms, can serve as a generalized reaction coordinate for various structural transitions in materials. By harmonically coupling the three components of each atom's NNOAD with three extra dynamic variables assigned to the same atom, we propose a set of dynamic equations (Eq. (2)) for the extended system with $3N$ additional dynamic variables. Atomic motions in real space follow Nosé-Hoover dynamics, while the extra dynamic variables evolve according to Langevin dynamics.

Through careful analysis of the SAMD equations and trial simulations of a benchmark problem, we determined the appropriate parameter ranges for the equations of the method. Case studies — including H segregation at GB in Al, shear response of GB in Al, tensile loading of α-Fe bicrystal, and Ag surface diffusion — demonstrate that SAMD provides effective accelerated simulation and consistent prediction of the various microstructure evolution behaviors compared to conventional MD simulations. The speedup ratio can reach around 8 orders of magnitude in certain cases. These results suggest that SAMD can serve as an accelerated MD method for atomistic simulation of a wide range of materials science problems concerning microstructure evolution which are beyond the reach of conventional MD.

We attribute the effectiveness and consistency in accelerated simulation and prediction of microstructure evolution using the SAMD method to the dynamic features of its governing equations. A comprehensive theoretical analysis and rigorous proof of this argument, however, remains elusive and warrant further investigation.

## CRediT authorship contribution statement

**Liang Wan:** Writing – original draft, Writing – review & editing, Methodology, Software, Investigation, Data curation, Conceptualization. **Qingsong Mei:** Conceptualization, Supervision, Resources. **Haowen Liu:** Software, Conceptualization, Project administration, Funding acquisition. **Huafeng Zhang:** Conceptualization, Supervision, Resources. **Jun-Ping Du:** Methodology, Conceptualization. **Shigenobu Ogata:** Writing – review & editing, Methodology, Supervision. **Wen Tong Geng:** Writing – review & editing, Conceptualization, Supervision.

## Declaration of Competing Interest

The authors declare that they have no known competing financial interests or personal relationships that could have appeared to influence the work reported in this paper.

## Data availability

Data will be made available on request.

## Acknowledgement

This work was supported by the National Key Research and Development Program of China (No. 2022YFB3707501) and the National Natural Science Foundation of China (grant 52175358). The numerical calculations in this paper have been done on the supercomputing system in the Supercomputing Centre of Wuhan University.

## Supplementary material

Supplementary movies S1-S5 associated with this article can be found in the online version, at doi:10.1016/j.cpc.2025.109943.

## Appendix A. Implementation of the SAMD method

A compressed archive can be downloaded at the online repository link https://gitee.com/lwan5/lammps-samd. In this compressed archive, a patch file implementing SAMD functionality in official LAMMPS releases, a set of LAMMPS input scripts for reproducing the results in Fig. 3, and a documentation detailing file usage and implementation procedures are provided.

## References

[1] S. Yip, *Handbook of Materials Modeling* (Springer, Dordrecht, 2005).

[2] W. Andreoni and S. Yip, *Handbook of Materials Modeling* (Springer, Switzerland, 2020), 2nd edn.

[3] M. Tuckerman, *Statistical Mechanics: Theory and Molecular Simulation* (Oxford University Press, Oxford, 2010).

[4] M. P. Allen and D. J. Tildesley, *Computer Simulation of Liquids* (Oxford University Press, Oxford, 2017).

[5] G. Ciccotti, C. Dellago, M. Ferrario, E. Hernández, and M. Tuckerman, *Molecular Simulations: Past, Present, and Future (a Topical Issue in Epjb)*, Eur. Phys. J. B **95**, 1 (2022).

[6] A. F. Voter, *Hyperdynamics: Accelerated Molecular Dynamics of Infrequent Events*, Phys. Rev. Lett. **78**, 3908 (1997).

[7] A. F. Voter, *Parallel Replica Method for Dynamics of Infrequent Events*, Phys Rev B Condens Matter **57**, R13985 (1998).

[8] M. R. Sorensen and A. F. J. Voter, *Temperature-Accelerated Dynamics for Simulation of Infrequent Events*, J. Chem. Phys. **112**, 9599 (2000).

[9] A. Laio and M. Parrinello, *Escaping Free-Energy Minima*, Proc. Natl. Acad. Sci. U.S.A. **99**, 12562 (2002).

[10] R. A. Miron and K. A. Fichthorn, *Accelerated Molecular Dynamics with the Bond-Boost Method*, J. Chem. Phys. **119**, 6210 (2003).

[11] A. Barducci, G. Bussi, and M. Parrinello, *Well-Tempered Metadynamics: A Smoothly Converging and Tunable Free-Energy Method*, Phys. Rev. Lett. **100**, 020603 (2008).

[12] A. Laio and F. L. Gervasio, *Metadynamics: A Method to Simulate Rare Events and Reconstruct the Free Energy in Biophysics, Chemistry and Material Science*, Rep. Prog. Phys. **71**, 126601 (2008).

[13] S. Hara and J. Li, *Adaptive Strain-Boost Hyperdynamics Simulations of Stress-Driven Atomic Processes*, Phys. Rev. B **82**, 184114 (2010).

[14] A. Ishii, S. Ogata, H. Kimizuka, and J. Li, *Adaptive-Boost Molecular Dynamics Simulation of Carbon Diffusion in Iron*, Phys. Rev. B **85**, 064303 (2012).

[15] S. Y. Kim, D. Perez, and A. F. Voter, *Local Hyperdynamics*, J. Chem. Phys. **139**, 144110 (2013).

[16] P. Tiwary and A. v. d. Walle, *A Review of Enhanced Sampling Approaches for Accelerated Molecular Dynamics*, in *Multiscale Materials Modeling for Nanomechanics*, edited by C. R. Weinberger, and G. J. Tucker (Springer, Switzerland, 2016), pp. 195-221.

[17] G. Fiorin and M. L. Klein, *Using Collective Variables to Drive Molecular Dynamics Simulations*, Mol. Phys. **111**, 3345 (2013).

[18] G. Bussi, A. Laio, and P. Tiwary, *Metadynamics: A Unified Framework for Accelerating Rare Events and Sampling Thermodynamics and Kinetics*, in *Handbook of Materials Modeling: Methods: Theory and Modeling*, edited by W. Andreoni, and S. Yip (Springer, Switzerland, 2020), pp. 565-595.

[19] M. Chen, *Collective Variable-Based Enhanced Sampling and Machine Learning*, Eur. Phys. J. B **94**, 1 (2021).

[20] J. Rogal, *Reaction Coordinates in Complex Systems-a Perspective*, Eur. Phys. J. B **94**, 1 (2021).

[21] J. Zhang and M. Chen, *Unfolding Hidden Barriers by Active Enhanced Sampling*, Phys. Rev. Lett. **121**, 010601 (2018).

[22] L. Maragliano and E. Vanden-Eijnden, *A Temperature Accelerated Method for*

*Sampling Free Energy and Determining Reaction Pathways in Rare Events Simulations*, Chem. Phys. Lett. **426**, 168 (2006).

[23] L. Rosso, P. Mináry, Z. Zhu, and M. E. Tuckerman, *On the Use of the Adiabatic Molecular Dynamics Technique in the Calculation of Free Energy Profiles*, The Journal of chemical physics **116**, 4389 (2002).

[24] J. B. Abrams and M. E. Tuckerman, *Efficient and Direct Generation of Multidimensional Free Energy Surfaces Via Adiabatic Dynamics without Coordinate Transformations*, The Journal of Physical Chemistry B **112**, 15742 (2008).

[25] T.-Q. Yu, P.-Y. Chen, M. Chen, A. Samanta, E. Vanden-Eijnden, and M. Tuckerman, *Order-Parameter-Aided Temperature-Accelerated Sampling for the Exploration of Crystal Polymorphism and Solid-Liquid Phase Transitions*, J. Chem. Phys. **140**, 214109 (2014).

[26] J. Rogal, E. Schneider, and M. E. Tuckerman, *Neural-Network-Based Path Collective Variables for Enhanced Sampling of Phase Transformations*, Phys. Rev. Lett. **123**, 245701 (2019).

[27] H. C. Andersen, *Molecular Dynamics Simulations at Constant Pressure and/or Temperature*, J. Chem. Phys. **72**, 2384 (1980).

[28] M. Parrinello and A. Rahman, *Polymorphic Transitions in Single-Crystals - a New Molecular-Dynamics Method*, J. Appl. Phys. **52**, 7182 (1981).

[29] S. Nose, *A Unified Formulation of the Constant Temperature Molecular-Dynamics Methods*, J. Chem. Phys. **81**, 511 (1984).

[30] S. Nose, *A Molecular Dynamics Method for Simulations in the Canonical Ensemble*, Mol. Phys. **52**, 255 (1984).

[31] W. G. Hoover, *Canonical Dynamics: Equilibrium Phase-Space Distributions*, Phys. Rev. A **31**, 1695 (1985).

[32] W. G. Hoover, *Constant-Pressure Equations of Motion*, Phys. Rev. A **34**, 2499 (1986).

[33] J. W. Christian, *The Theory of Transformations in Metals and Alloys* (Pergamon, Oxford, 2002).

[34] R. C. Pond and S. Celotto, *Special Interfaces: Military Transformations*, Int. Mater. Rev. **48**, 225 (2003).

[35] P. M. Anderson, J. P. Hirth, and J. Lothe, *Theory of Dislocations* (Cambridge University Press, Cambridge, 2017).

[36] H. Van Swygenhoven and P. M. Derlet, *Grain-Boundary Sliding in Nanocrystalline Fcc Metals*, Phys. Rev. B **64**, 224105 (2001).

[37] B. Li and E. Ma, *Atomic Shuffling Dominated Mechanism for Deformation Twinning in Magnesium*, Phys. Rev. Lett. **103**, 035503 (2009).

[38] B.-Y. Liu *et al.*, *Twinning-Like Lattice Reorientation without a Crystallographic Twinning Plane*, Nat. Commun. **5**, 3297 (2014).

[39] A. Ishii, J. Li, and S. Ogata, *Shuffling-Controlled Versus Strain-Controlled Deformation Twinning: The Case for Hcp Mg Twin Nucleation*, Int. J. Plast. **82**, 32 (2016).

[40] L.-L. Niu, Y. Zhang, X. Shu, F. Gao, S. Jin, H.-B. Zhou, and G.-H. Lu, *Shear-Coupled Grain Boundary Migration Assisted by Unusual Atomic Shuffling*, Sci. Rep. **6**, 1 (2016).

[41] L. Wan, A. Ishii, J.-P. Du, W.-Z. Han, Q. Mei, and S. Ogata, *Atomistic Modeling Study of a Strain-Free Stress Driven Grain Boundary Migration Mechanism*, Scripta Mater. **134**, 52 (2017).

[42] M. L. Falk and J. S. Langer, *Dynamics of Viscoplastic Deformation in Amorphous Solids*, Phys. Rev. E **57**, 7192 (1998).

[43] M. L. Falk, *Molecular-Dynamics Study of Ductile and Brittle Fracture in Model Noncrystalline Solids*, Phys. Rev. B **60**, 7062 (1999).

[44] W. F. van Gunsteren and H. J. C. Berendsen, *Algorithms for Brownian Dynamics*, Mol. Phys. **45**, 637 (1982).

[45] D. S. Lemons and A. Gythiel, *Paul Langevin's 1908 Paper "on the Theory of Brownian Motion"["Sur La Théorie Du Mouvement Brownien," Cr Acad. Sci.(Paris) 146, 530–533 (1908)]*, Am. J. Phys. **65**, 1079 (1997).

[46] L. Maragliano and E. Vanden-Eijnden, *Single-Sweep Methods for Free Energy Calculations*, The Journal of Chemical Physics **128**, 184110 (2008).

[47] A. P. Thompson et al., *LAMMPS - a flexible simulation tool for particle-based materials modeling at the atomic, meso, and continuum scales*, Comp. Phys. Comm. **271**, 10817 (2022).

[48] LAMMPS official website: https://www.lammps.org/

[49] D. Faken and H. Jonsson, *Systematic Analysis of Local Atomic Structure Combined with 3d Computer Graphics*, Comput. Mater. Sci. **2**, 279 (1994).

[50] C. S. Becquart, J. M. Raulot, G. Bencteux, C. Domain, M. Perez, S. Garruchet, and H. Nguyen, *Atomistic Modeling of an Fe System with a Small Concentration of C*, Comput. Mater. Sci. **40**, 119 (2007).

[51] R. G. Veiga, M. Perez, C. S. Becquart, and C. Domain, *Atomistic Modeling of Carbon Cottrell Atmospheres in Bcc Iron*, J. Phys.: Condens. Matter **25**, 025401 (2013).

[52] H. Jónsson, G. Mills, and K. W. Jacobsen, in *Classical and Quantum Dynamics in Condensed Phase Simulations* (1998), pp. 385-404.

[53] G. Henkelman, B. P. Uberuaga, and H. Jonsson, *A Climbing Image Nudged Elastic Band Method for Finding Saddle Points and Minimum Energy Paths*, J. Chem. Phys. **113**, 9901 (2000).

[54] *Diffusion in Solid Metals and Alloys*, in *Defect and Diffusion Forum Vol. 3* (Trans Tech Publications Ltd, 1969), pp. 119-182.

[55] P. Shewmon, *Diffusion in Solids* (Springer, Switzerland, 2016).

[56] S. M. Allen, R. W. Balluffi, and W. C. Carter, *Kinetics of Materials* (John Wiley & Sons, New Jersey, 2005).

[57] R. P. Gangloff and B. P. Somerday, *Gaseous Hydrogen Embrittlement of Materials in Energy Technologies* (Woodhead Publishing Limited, Cambridge, 2012).

[58] M. Nagumo, *Fundamentals of Hydrogen Embrittlement* (Springer, Singapore, 2016).

[59] L. Wan, W. T. Geng, A. Ishii, J.-P. Du, Q. Mei, N. Ishikawa, H. Kimizuka, and S. Ogata, *Hydrogen Embrittlement Controlled by Reaction of Dislocation with Grain Boundary in Alpha-Iron*, Int. J. Plast. **112**, 206 (2019).

[60] D.-G. Xie, L. Wan, and Z.-W. Shan, *Hydrogen Enhanced Cracking Via Dynamic Formation of Grain Boundary inside Aluminium Crystal*, Corros. Sci. **183**, 109307 (2021).

[61] F. Apostol and Y. Mishin, *Angular-Dependent Interatomic Potential for the Aluminum-Hydrogen System*, Phys. Rev. B **82**, 144115 (2010).

[62] Y. Mishin, D. Farkas, M. J. Mehl, and D. A. Papaconstantopoulos, *Interatomic Potentials for Monoatomic Metals from Experimental Data and Ab Initio Calculations*, Phys. Rev. B **59**, 3393 (1999).

[63] L. Wan and S. Wang, *Shear Response of the Σ11, ⟨1 1 0⟩{1 3 1} Symmetric Tilt Grain Boundary Studied by Molecular Dynamics*, Modell. Simul. Mater. Sci. Eng. **17**, 045008

(2009).

[64] W. Shinoda, M. Shiga, and M. Mikami, *Rapid Estimation of Elastic Constants by Molecular Dynamics Simulation under Constant Stress*, Phys. Rev. B **69**, 134103 (2004).

[65] T. Zhu, J. Li, A. Samanta, A. Leach, and K. Gall, *Temperature and Strain-Rate Dependence of Surface Dislocation Nucleation*, Phys. Rev. Lett. **100**, 025502 (2008)

[66] M. I. Mendelev, S. Han, D. J. Srolovitz, G. J. Ackland, D. Y. Sun, and M. Asta, *Development of New Interatomic Potentials Appropriate for Crystalline and Liquid Iron*, Philos. Mag. **83**, 3977 (2003).

[67] G. J. Ackland, M. I. Mendelev, D. J. Srolovitz, S. Han, and A. V. Barashev, *Development of an Interatomic Potential for Phosphorus Impurities in a-Iron*, J. Phys.: Condens. Matter **16**, S2629 (2004).

[68] P. Williams, Y. Mishin, and J. Hamilton, *An Embedded-Atom Potential for the Cu–Ag System*, Modell. Simul. Mater. Sci. Eng. **14**, 817 (2006).

[69] H. A. Kramers, *Brownian Motion in a Field of Force and the Diffusion Model of Chemical Reactions*, Physica **7**, 284 (1940).

[70] P. Hänggi, P. Talkner, and M. Borkovec, *Reaction-Rate Theory: Fifty Years after Kramers*, Rev. Mod. Phys. **62**, 251 (1990).

[71] R. Zwanzig, *Nonequilibrium Statistical Mechanics* (Oxford University Press, New York, 2001).

[72] L. Peliti and S. Pigolotti, *Stochastic Thermodynamics: An Introduction* (Princeton University Press, New Jersey, 2021).

[73] E. Witten, *String theory dynamics in various dimensions*, Nucl. Phys. B **443**, 85 (1995).

[74] S. T. Yau and S. J. Nadis, *The shape of inner space: String theory and the geometry of the universe's hidden dimensions* (Il Saggiatore, New York, 2010).

[75] A. N. Schellekens, *Life at the interface of particle physics and string theory*, Rev. Mod. Phys. **85**, 1491 (2013).